\documentclass{article}
\usepackage{amsmath,graphicx,subfig}
\usepackage{caption}
\usepackage{amssymb,amsbsy}
\author{L. Albano Farias and J. Stephany}
\title{Optimization of the transmission of observable expectation values and observable
statistics in Continuous Variable Teleportation}

\begin{document}
\begin{titlepage}

\hfill{Preprint {\bf SB/F/383-10}} \hrule \vskip .5cm

\begin{center}{\large{\bf Optimization of the transmission of observable expectation values \\
and observable statistics in Continuous Variable Teleportation }}
\vskip 0.5cm
 L. Albano Farias and J. Stephany
\\
{\it Departamento de F\'{\i}sica, Secci\'{o}n de Fen\'{o}menos \'{O}pticos, \\
Universidad Sim\'{o}n Bol\'{\i}var,\\
Apartado Postal 89000, Caracas 1080A, Venezuela}
\vskip 0.3mm

{\it e-mail:lalbano@fis.usb.ve,stephany@usb.ve}
\end{center}

\begin{abstract}

We analyze the statistics of observables  in continuous variable (\textbf{CV}) quantum teleportation in the formalism of the characteristic function. We derive expressions for average values of output state observables in particular cumulants which are additive  in terms of the input state and  the resource of teleportation..

Working with  a general class of teleportation resources, the Squeezed Bell-like states, which may  be optimized in a free parameter for better teleportation performance~\cite{TelepNonGauss} we discuss the relation between  resources optimal for fidelity and for different observable averages. We obtain the values of the free parameter of the Squeezed Bell-like states which optimize the  central  momenta and cumulants up to fourth order. For the cumulants the distortion between in and out states due to teleportation  depends only on the resource. We obtain optimal parameters $\Delta_{(2)}^{opt}$ and $\Delta_{(4)}^{opt}$ for the second and fourth order cumulants which do not depend on the squeezing of the resource. The second order central momenta which is equal to the second order cumulants and the photon number average are also optimized by the resource with $\Delta_{(2)}^{opt}$.

We show that the optimal fidelity resource which has been found in reference~\cite{TelepNonGauss}
to depend also on the characteristics of input tends for high squeezing to the resource which
optimizes the second order momenta. A similar behavior is obtained for the resource which optimizes
the photon statistics which is treated here using the sum of the squared differences in photon
probabilities of input and output states as the distortion measure. This is interpreted naturally
to mean that the distortions associated to second order momenta dominates the behavior of the
output state for large squeezing of the resource. Optimal fidelity resources and optimal photon statistics resources are compared and is shown that for mixtures of Fock states  both resources are equivalent.
\end{abstract}
\vskip 0.3cm
\hrule
\bigskip
\centerline{\bf UNIVERSIDAD SIMON BOLIVAR}
\vfill
\end{titlepage}

\section{Introduction}

Continuous variable \textbf{CV} quantum teleportation ~\cite{BraunsteinKimble}, depends on the use
of a two mode quantum entangled resource and in general allows only the reconstruction of an
imperfect output state.  It has been shown that the success probability of teleportation can be
increased by using entangled non-Gaussian resources which can be chosen to improve the efficiency
of the protocol ~\cite{Opatrny00,Olivares03,Kitagawa06}. In particular for a specific family of
two-mode non-Gaussian states, the squeezed Bell-like states which include photon-added and photon-
subtracted  states it has been shown that fidelity between input and output states can be improved
by carefully choosing the parameters of the resource~\cite{TelepNonGauss,re:LorenzoTesis}. For high
squeezing squeezed Bell-like states tend to $EPR$ states and teleportation becomes perfect. Other
classes of non-Gaussian resources, such as two-mode squeezed symmetric superpositions of Fock
states and of squeezed cat-like states, with good performance in the teleportation of single-mode
input states were considered in ~\cite{re:LorenzoTesis,TelepNonGauss2}.

In general, teleportation success is increased by maximizing a functional of the density matrices
of the output and the input states, such as the fidelity~\cite{re:BraunsteinFuchsKimble}, over
modifications of the resource and the teleportation protocol. A more general approach may be
considered since on one side fidelity is not directly observable except for input states that are
eigenstates of an observable quantity and in the other any verifying process for the output of
teleportation will consist in a series of measurements on the output and the input states.
Therefore, it is worthwhile to study how to optimize the transmission of observable averages and/or
observable statistics in an output state and to compare such optimal transmission situations with
optimal fidelity situations. For specific experimental situations, the preservation of observable
averages and associated variances may be in principle of more interest than the overall fidelity.
Analysis of the optimal teleportation of average values associated with non-classicality for
conditionally produced resources has been done in reference~\cite{re:MazeOrszag}, using the
transfer operator description of \textbf{CV} teleportation~\cite{re:TransferOpHofmann}.
Teleportation of the oscillations of photon statistics of squeezed, non-classical states has been
done in~\cite{re:MundarainOrszag}

In section~(\ref{AverOut}) the  characteristic function description of
teleportation~\cite{re:MarianMarian,TelepNonGauss,re:LorenzoTesis} is used to obtain averages of
observables for the output in terms of contributions by the resource and the input. The differences
between cumulants and raw momenta are stressed. In section~(\ref{OptAvEx}) the preservation  in
teleportation of cumulants and momenta is optimized for selected input states over the Squeezed
Bell-like class of non-Gaussian resources. These optimal preservation resources are compared with
optimal fidelity resources. In sections~(\ref{Dfunc}) and~(\ref{DFuncSamples}) we use the
functional $\mathcal{D}_N$ of the quadratic  deviations of photon probabilities to study the
distortion of photon statistics through teleportation. We compare  $\mathcal{D}_N$  with Frobenius distance and fidelity. Finally in section~(\ref{Discuss}) we present our conclusion.

\section{Averages of observables for the teleportation output}\label{AverOut}

\subsection{The characteristic function description of teleportation}\label{AverOut:charf}
Quantum Teleportation may be described in a very convenient form in terms of characteristic
functions of the input, the output, and the resource states. For a quantum system with complex
phase space variable $\alpha=x\,+\,\imath\,p$ the characteristic functions are  Fourier transforms
of phase space pseudo-distribution functions such as  Wigner's, $P(\alpha)$ and $Q(\alpha)$
functions. They are  generating functions for the raw moments associated with their corresponding
pseudo-probability function~\cite{DistFuncWigner}. $P(\alpha)$~\cite{re:MarianMarian}
characteristic functions are associated with averages of normally ordered products $\hat{a}^{\dag\,
n}\hat{a}^{m}$  and  Wigner's characteristic functions~\cite{re:LorenzoTesis} are associated with
averages of symmetric ordered products. If $\xi\equiv\,w\,+\,\imath\,z$ is the conjugate coordinate
to the phase space coordinate $\alpha$ they are related by
\begin{equation}\label{eq:CharFuncOrder}
   \chi(\xi)^{(s)}=\,e^{s\,|\xi|^{2}/2}\chi(\xi)^{(0)}\ \ ,
\end{equation}
with the symmetric ordering Wigner's characteristic ($s=0$) function being
$\chi(\xi)=\chi(\xi)^{(0)}$ and the $P(\alpha)$ characteristic function ($s=1$) for normal ordering
being $\chi(\xi)^{(1)}$.

For quantum teleportation in \textbf{CV}, using a symmetric Beam Splitter that  mixes the $A$ and
$in$ states (T=$\cos(\pi/4)^2$),  with a measurement gain  $g$ for the homodyne measurement and for
the second partner (\textbf{Bob}) correction, any one of the characteristic functions for the
output state can be written in terms of the characteristic function of the input state as,
\begin{equation}\label{eq:CVTelepoutoneg}
\chi_{out}(\xi_{B}) =\chi_{AB}(g \xi_{B}^{*}; \xi_{B}) \: \chi_{in}(g \xi_{B})\ \ .
\end{equation}
We use both $P(\alpha)$ and Wigner's characteristic
functions in what follows as appropriate and take $g=1$ for the purposes of this work.

The output characteristic function in equation~(\ref{eq:CVTelepoutoneg}) is a product of the
characteristic function of the input state $\chi_{in}(g\,\xi)$ and the characteristic function of
the resource $\chi_{AB}(g\,\xi^{*},\xi)$ evaluated at a certain point in the two-mode phase space.
It is called the \textit{Transfer Function} of the setup, and has the one-mode phase
space coordinate $\xi$ as argument.
For a Gaussian two mode resource state,  this has been shown to correspond to a one-mode Gaussian characteristic function
~\cite{re:MarianMarianNewLook}, amounting in the output state to a distorting Gaussian noise.

The distortion of the teleported state in \textbf{CV} depends on the transfer function. The
transfer function of an state approaching the \textbf{EPR} state is nearly constant, being close to
the value $1$ at each point of the one-mode phase space. Thus the limit EPR state is not
square-integrable~\cite{re:LorenzoTesis}. From equation~(\ref{eq:CVTelepoutoneg}) it is seen that
with a state approximating an \textbf{EPR} state  as a resource the output state is nearly
identical to the input state.

\subsection{Averages of operators for the output of teleportation}\label{AverOut:avout}

For a particular ordering $(s)$ of the creation and annihilation operators $\hat{a}^{\dag}$ and $\hat{a}$,

\begin{equation}\label{eq:CharAvgCreAnn}
\left \langle \hat{a}^{\dag \; n}\hat{a}^{m} \right \rangle^{(s)} \;=\; \left.
\frac{\partial^{n+m}\chi_{out}^{(s)}(\xi)}{\partial\,\xi^{n}\;\partial\,\xi^{*\,m}}\right|_{\xi=\xi^{*}=0}\ \ .
\end{equation}

The averages of $\hat{x}=(\hat{a}+\hat{a}^{\dag})/2$ and $\hat{p}=(\hat{a}-\hat{a}^{\dag})/2\,i$
position and momentum operators in  \textbf{CV} can be obtained using Wigner's characteristic
function, given that products $\hat{x}^{n}\hat{p}^{m}$ are symmetrically ordered. Thus, with
$\xi=w\;+\;i\,z$ we have,

\begin{equation}\label{eq:CharAvgXP}
\left \langle \hat{x}^{\dag \; n}\hat{p}^{m} \right \rangle \;=\; \frac{1}{(i)^{n+m}}\left.
\frac{\partial^{n+m}\chi_{out}(\xi)}{\partial\,z^{n} \;\partial\,w^{m}}\right|_{w=z=0}\ .
\end{equation}

Now, it is
straightforward to calculate the expectation values associated with the teleportation output, in
terms of the contribution of averages associated with the input state, and with the derivatives,
evaluated at $\xi=\xi^{*}=0$ of the transfer function. Though the transfer function may not
necessarily be a proper characteristic function of a one mode state, for convenience and economy of notation
their derivatives evaluated at $\xi=\xi^{*}=0$ will be denoted as operator averages. Given
\begin{eqnarray}
 \left \langle \hat{a}^{\dag \; n}\hat{a}^{m} \right \rangle_{in}&=&\left.\frac{1}{(i)^{n+m}}\;\frac{\partial^{n+m}\chi_{in}(\xi)}{\partial\,\xi^{n}\;\partial\,\xi^{*\,m}}\right|_{\xi=\xi^{*}=0} \label{eq:AvgCharCAIn}\ \ , \\
\left \langle \hat{x}^{\; n}\hat{p}^{m} \right \rangle_{in}&=&\left.\frac{\partial^{n+m}\chi_{in}(\xi)}{\partial\,z^{n}\;\partial\,w^{m}}\right|_{w=z=0} \label{eq:AvgCharXPIn}\ \ ,\\
\left \langle \hat{a}^{\dag \; n}\hat{a}^{m} \right \rangle_{\widetilde{AB}} &\equiv& \left.
\frac{\partial^{n+m}\chi_{AB}(g\xi^{*},\xi)}{\partial\,\xi^{n}\;\partial\,\xi^{*\,m}}\right|_{\xi=\xi^{*}=0} \label{AvgCharCAAB}\ \ , \\
\left \langle \hat{x}^{n}\hat{p}^{m} \right \rangle_{\widetilde{AB}} &\equiv&
\left.\frac{1}{(i)^{n+m}}\;
\frac{\partial^{n+m}\chi_{AB}(g\xi^{*},\xi)}{\partial\,z^{n}\;\partial\,w^{*\,m}}\right|_{w=z=0}\ \ ,
\label{eq:AvgCharXPAB}
\end{eqnarray}
the expectation values for $\hat{a}^{\dag \; n}\hat{a}^{m}$ associated with the output state (see
equations~(\ref{eq:CVTelepoutoneg}),~(\ref{eq:CharAvgCreAnn})) are written as,
\begin{equation}\label{eq:CharAvgCreAnnOut1}
\left \langle \hat{a}^{\dag \; n}\hat{a}^{m} \right \rangle_{out} \;=\; \sum_{i=0,j=0}^{n,m}
\left(\begin{array}{c}n\\i\end{array}\right) \left(\begin{array}{c}m\\j\end{array}\right)
g^{i+j}\left\langle \hat{a}^{\dag \; i}\hat{a}^{j} \right \rangle_{in}\;\left \langle \hat{a}^{\dag
\; n-i}\hat{a}^{m-j} \right \rangle_{\widetilde{AB}}\ \ .
\end{equation}
Likewise, for the $\hat{x}^{n}\hat{p}^{m}$ averages, we have
\begin{equation}\label{eq:CharAvgXPOut1}
\left \langle \hat{x}^{n}\hat{p}^{m} \right \rangle_{out} \;=\; \sum_{i=0,j=0}^{n,m}
\left(\begin{array}{c}n\\i\end{array}\right) \left(\begin{array}{c}m\\j\end{array}\right)
g^{i+j}\left\langle \hat{x}^{i}\hat{p}^{j} \right \rangle_{in}\;\left\langle
\hat{x}^{n-i}\hat{p}^{m-j} \right\rangle_{\widetilde{AB}}\ \ .
\end{equation}

The expectation values for the output state are expressed as a sum of products of "averages" of
operators for the input state and the transfer function.

For the average of $\langle\, \hat{a}^{\dag \; n}\hat{a}^{m}\, \rangle_{out}$, we have the
terms $g^{n+m}\;\langle\, \hat{a}^{\dag \; n}\hat{a}^{m}\,\rangle_{in}$ of the input state and
the average $\langle\, \hat{a}^{\dag \; n}\hat{a}^{m}\, \rangle_{\widetilde{AB}}$ as $n+m$ order
terms. The expectation value of $\hat{a}^{\dag \; n}\hat{a}^{m}$ for the output state can be
expressed as the equivalent average for the input state (multiplied by the power of the gain
$g^{n+m}$), plus terms that cause distortion of the input expectation value depending on products
of lesser-order "averages" of the input state and the transfer function, plus a term depending on $\langle\,\hat{a}^{\dag \; n}\hat{a}^{m}\, \rangle_{\widetilde{AB}}$.

The $\langle\,\hat{x}^{n}\hat{p}^{m}\, \rangle_{out}$ expectation value of the output state has a
similar form. It can be described as the expectation value of the input state (multiplied by
$g^{n+m}$), $g^{n+m}\left\langle \hat{x}^{n}\hat{p}^{m} \right \rangle_{in}$, plus distorting terms
depending on products of lesser-order "averages" of the input and transfer function, plus the
$\langle\,\hat{x}^{n}\hat{p}^{m}\, \rangle_{\widetilde{AB}}$ "average" for the transfer function.

A related, resource-specific  analysis for Gaussian resources and photon subtracted resources of the observable expectation values for the output state of teleportation can be found in
reference~\cite{re:MazeOrszag}.

\subsection{The transfer function of a two-mode squeezed state}\label{Averout:transfertwosq}

Wigner's characteristic function of a two-mode resource state prepared from a previous symmetric state of modes  \textbf{A} and \textbf{B} by two-mode squeezing operation with  phase of $\pi$,  can be
written as $\chi_{AB}(\xi_{A}';\xi_{B}')$, with transformed coordinates
$\xi_{A/B}'(\xi_{A};\xi_{B})=\xi_{A/B}\cosh(r)-\xi_{B/A}^{*}\sinh(r)$. ~\cite{Barnett}.

Therefore, the transfer function for such a resource has arguments
$\xi_{A}'(\xi^{*};\xi)=\xi^{*}\,e^{-r}$ and $\xi_{B}'(\xi^{*};\xi)=\xi\,e^{-r}$. It can then be
written as
\begin{equation}\label{eq:transfuncsq}
    \chi_{AB}(\xi^{*}\,e^{-r};\xi\,e^{-r})\ \ .
\end{equation}

The states  prepared by two-mode squeezing have Wigner's
characteristic functions with the argument $|\xi|^{2}$. These Wigner's characteristic functions
always have a multiplying factor of the form $e^{-|\xi|^{2}/2}$, which ensures that they are
square-integrable, even if they are very non-Gaussian. A transfer function of the form outlined above
can be written as
\begin{equation}\label{eq:transfuncsq2}
    \chi_{AB}(|\xi|^{2}\,e^{-2r})
\end{equation}
having an envelope of the form $e^{-|\xi|^{2}e^{-2r}}$. The transfer function, being the
characteristic function of a two mode state evaluated at a particular point in two-mode conjugate
phase space, satisfies that $\chi_{AB}(0;0)=1$~\cite{Barnett}.

The transfer function is  real and symmetric in the conjugate phase space defined by
$\xi=w\,+\,iz$, as exponential and polynomial factors with argument $|\xi|^{2}e^{-2r}$ are
symmetric around $\xi=0$. While not necessarily Gaussian, the transfer function is
square-integrable and has a maximum at $\xi=0$ with value $1$; given that $e^{-|\xi|^{2}e^{-2r}}$
decreases faster than any polynomial factor in the transfer function around $\xi=0$. This maximum
is more salient for small squeezing $r$. For large $r$, at and near $\xi=0$ the transfer function
is approximately constant.

Therefore, the first derivatives of this transfer function, at $\xi=0$ are $0$. Second order
derivatives are negative or $0$ at $\xi=0$. Thus, for the the transfer function of a two-mode
squeezed state,
\begin{equation} \label{eq:1stordavtransfereqzero}
    \left \langle \hat{a}^{\dag} \right \rangle_{\widetilde{AB}}=\left \langle \hat{a} \right
    \rangle_{\widetilde{AB}}=\left \langle \hat{x} \right \rangle_{\widetilde{AB}}=\left \langle \hat{p} \right
    \rangle_{\widetilde{AB}}=0\ \ .
\end{equation}

\begin{figure}[tb]
\centering
%%----primera----
\subfloat[Squeezed Vacuum]{
\includegraphics[width=0.35\linewidth]{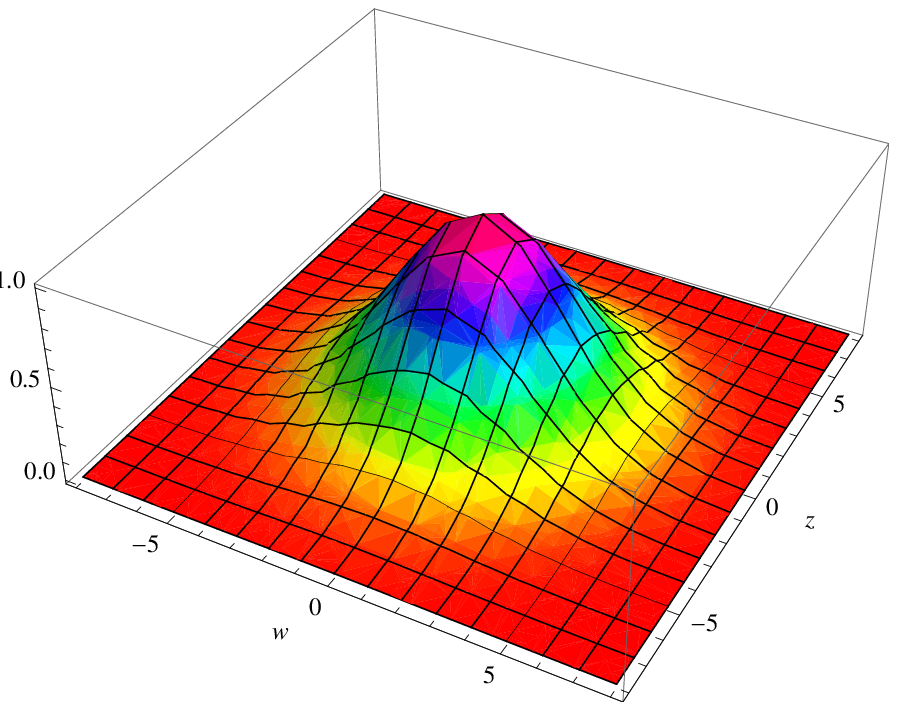}}
\hspace{0.1\linewidth}
%%----start of second subfigure----
\subfloat[Photon-Subtracted]{
\includegraphics[width=0.35\linewidth]{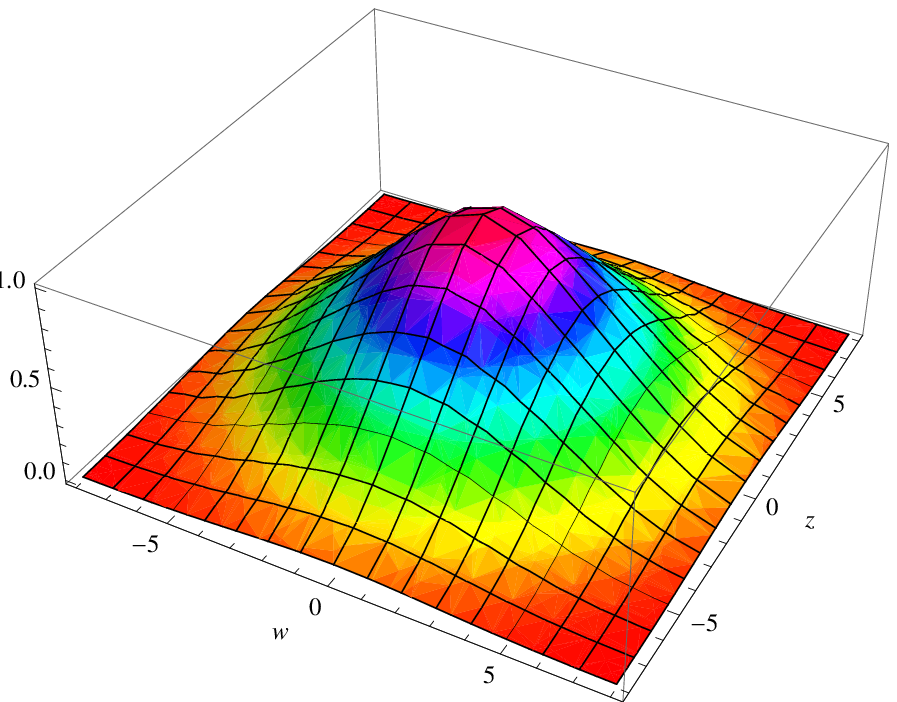}}\\
%%----start of third subfigure----
\subfloat[Photon-Added]{
\includegraphics[width=0.35\linewidth]{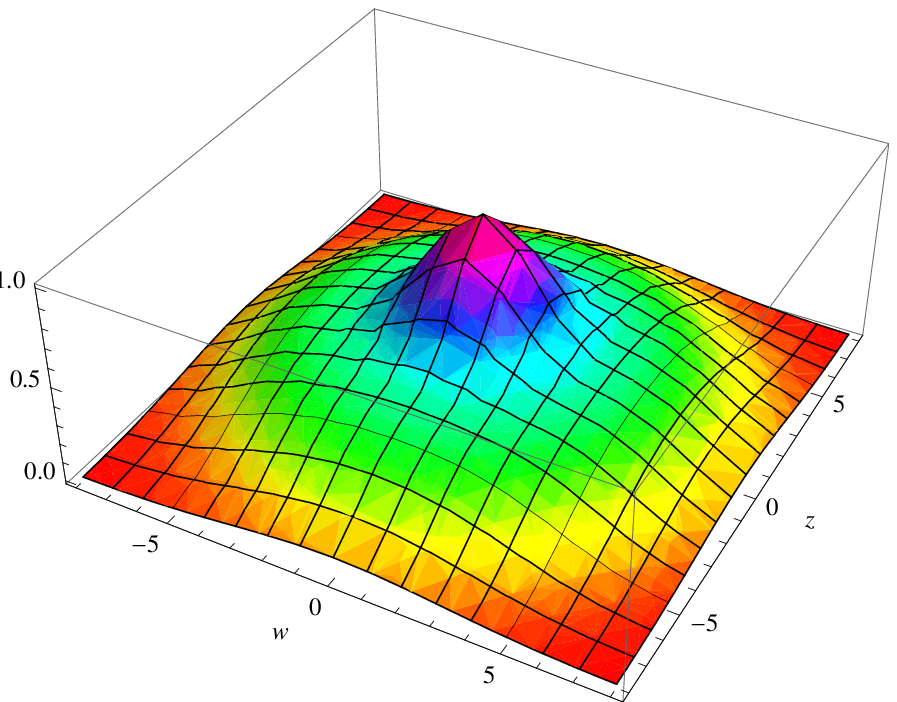}}
\hspace{0.1\linewidth}
%%----start of fourth subfigure----
\subfloat[Optimal for Coherent Input]{
\includegraphics[width=0.35\linewidth]{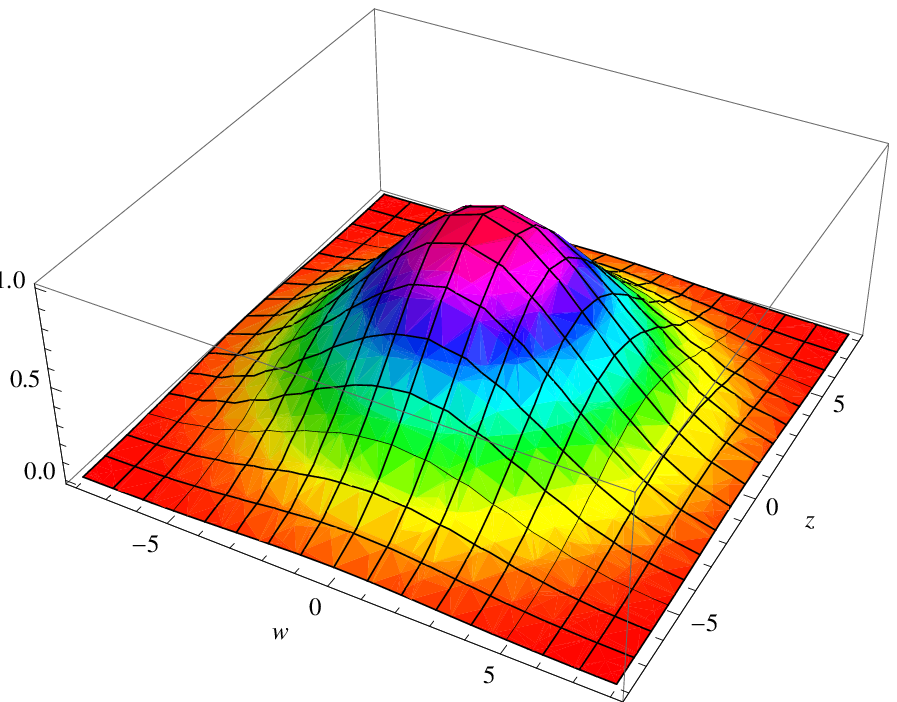}}
\caption{The transfer function $\chi(\xi^{*},\xi)$  with two-mode squeezing $r=1.25$ for: (a)
Squeezed Vacuum; (b) A photon-subtracted resource; (c) A photon-added resource; (d) A Squeezed
Bell-like resource optimized for the teleportation of a coherent input state}
\label{fig:transfuncexample} %% label for entire figure
\end{figure}

Teleportation  using a two-mode squeezed resource does not alter the position and momentum averages
$\left \langle \hat{x} \right \rangle_{in}$\,,\,$\left \langle \hat{p} \right \rangle_{in}$ of the
input state. To illustrate this discussion, figure~(\ref{fig:transfuncexample}) displays several
sample transfer functions associated with well-known teleportation resources.

The best resources for  teleportation  are those like the photon-subtracted two-mode squeezed
state~\cite{Opatrny00} and optimized Squeezed Bell-like state~\cite{TelepNonGauss}, which have a
smoother transfer function  around $\xi=0$.

\subsection{Output expectation values for relevant observables}

\subsubsection{Expectation value of the photon number}

The expectation value of the photon number operator $\hat{n}=\hat{a}^{\dagger}\hat{a}$ can be
calculated from equation~(\ref{eq:CharAvgCreAnnOut1}),
\begin{eqnarray}\label{eq:NOutAvg}
    \langle\,\hat{a}^{\dagger}\,\hat{a}\,\rangle_{out}&=&g^{2}\,\langle\,\hat{a}^{\dagger}\,\hat{a}\,\rangle_{in}
    \;+\;\langle\,\hat{a}^{\dagger}\,\hat{a}\,\rangle_{\widetilde{AB}}\;\nonumber\\
&+&\;g\,\langle\,\hat{a}\,\rangle_{in}\,\langle\,\hat{a}^{\dagger}\,\rangle_{\widetilde{AB}}
    \;+\;g\,\langle\,\hat{a}^{\dagger}\,\rangle_{in}\,\langle\,\hat{a}\,\rangle_{\widetilde{AB}}\ \ .
\end{eqnarray}

The expectation value for $\hat{n}$ of the input state is modified in the teleportation output, by
the addition of the "average number of photons" associated with the transfer function, and by two
cross-terms $g\,\langle\,\hat{a}^{\dagger}\,\rangle\langle\,\hat{a}\,\rangle_{\widetilde{AB}}$ and
$g\,\langle\,\hat{a}\,\rangle_{in}\langle\,\hat{a}^{\dagger}\,\rangle_{\widetilde{AB}}$. which are
equal to 0 (see equation~(\ref{eq:1stordavtransfereqzero})).

\subsubsection{Expectation value of Covariance matrix averages}

The transmission of squeezing by teleportation and  the success in  teleportation of a Gaussian
input state may be explored through the comparison of the expectation values of the covariance
matrix of the output and input states.

The second-order moments (in $\hat{x}$ and $\hat{p}$, and thus in $\hat{a}^{\dagger}$ and
$\hat{a}$) that make up the covariance matrix
\begin{eqnarray}
\langle\,\Delta
\hat{x}^{2}\,\rangle&=&\langle\,\hat{x}^{2}\,\rangle-\langle\hat{x}\rangle^{2}\label{eq:DeltaP2}\\
\langle\,\Delta \hat{p}^{2}\,\rangle&=&\langle\,\hat{p}^{2}\,\rangle-\langle\hat{p}\rangle^{2}\label{eq:DeltaX2}\\
\mathrm{Cov}(\hat{x},\hat{p})\,&=&\,\langle\,\hat{x}\hat{p}\,\rangle\,-\,\langle\,\hat{x}\,\rangle\,\langle\,\hat{p}\,\rangle
\label{eq:DeltaXP}
\end{eqnarray}
define  Gaussian states, up to
the average of $\hat{x}$ and $\hat{p}$ or equivalently, up to the application of a Glauber displacement operation
$e^{\hat{a}^{\dagger}\,\alpha\,-\,\hat{a}\,\alpha^{*}}$.

The quotient of the second order central moments
\begin{equation} \label{eq:Squeezing}
    \mathcal{S}\equiv\frac{\langle\,\Delta\,\hat{x}^{2}\,\rangle}{\langle\,\Delta\,\hat{p}^{2}\,\rangle}\ ,
\end{equation}
defines the squeezing $ \mathcal{S}$ of  one-mode states. For  $\mathcal{S}\neq 1$ states are
called squeezed. Squeezing of $\mathcal{S}=e^{-4\,r}$ is produced by applying the
transformation
$$e^{\frac{\zeta}{2}\,
\hat{a}^{\dagger\,2}-\frac{\zeta^{*}}{2} \,\hat{a}^{2}}\ \ ,$$ with a phase
$0$($\zeta=r\;\;r\in\,\mathbb{R}$) to a state that  initially is not squeezed.

According to equation~(\ref{eq:CharAvgXP}), the averages $\langle\,\hat{x}^{2}\,\rangle$ and
$\langle\hat{x}\rangle$ are given by
\begin{eqnarray}
\langle\hat{x}\rangle_{out}\,&=&\;g\,\langle\,\hat{x}\,\rangle_{in}+\langle\,\hat{x}\,\rangle_{\widetilde{AB}}
\label{eq:XAvgsqout}\\
\langle\hat{x}^{2}\rangle_{out}\,&=&\;g^{2}\,\langle\,\hat{x}^{2}\,\rangle_{in}\,+\,\langle\,\hat{x}^{2}\,\rangle_{\widetilde{AB}}\,+\,2\,g\,\langle\,\hat{x}\,\rangle_{in}\,\langle\,\hat{x}\,\rangle_{\widetilde{AB}}\ .
\label{eq:XXAvgsqout}
\end{eqnarray}

A straightforward calculation yields, for $\langle\,\Delta \hat{x}^{2}\,\rangle_{out}$,
\begin{eqnarray} \label{eq:DeltaXOut}
\langle\,\Delta
\hat{x}^{2}\,\rangle_{out}\,&=&\,g^{2}\,(\langle\,\hat{x}^{2}\,\rangle_{in}\,-\,\langle\,\hat{x}\,\rangle^{2}_{in})+\,(\langle\,\hat{x}^{2}\,\rangle_{\widetilde{AB}}\,-\,\langle\,\hat{x}\,\rangle^{2}_{\widetilde{AB}})
\,\nonumber\\
&=&\,g^{2}\langle\,\Delta \hat{x}^{2}\,\rangle_{in}\,+\,\langle\,\Delta
\hat{x}^{2}\,\rangle_{\widetilde{AB}}\ ,
\end{eqnarray}
where we have defined $\langle\,\Delta
\hat{x}^{2}\,\rangle_{\widetilde{AB}}\equiv\langle\,\hat{x}^{2}\,\rangle_{\widetilde{AB}}\,-\,\langle\,\hat{x}\,\rangle^{2}_{\widetilde{AB}}$.

Equation~(\ref{eq:DeltaXOut}) states that the difference between the variances of input and output
states in \textbf{CV} teleportation does not depend on the teleportation input but only on the
resource's transfer function. This may be traced to the fact that the second order central moment is identical to
the second order cumulant. The cumulant of order $n$ of a probability distribution that is the
convolution of two probability distributions, is equal to the sum of the cumulants of order $n$ of the
two probability distributions~\cite{Jaynes}. The characteristic function of the teleportation
output in equation~(\ref{eq:CVTelepoutoneg}) is a product of characteristic functions and  the
corresponding Wigner function is the convolution of Wigner functions of the input and of the
resource~\cite{BraunsteinKimble}. Below, we extend this observation to the third and fourth order
cumulants.

The minimization of the distortion of $\langle\,\Delta
\hat{x}^{2}\,\rangle_{out}$ for any input state can be carried out by minimizing the second
derivative
\begin{equation*}
    \langle\,\hat{x}^{2}\,\,\rangle_{\widetilde{AB}}=\left.\frac{\partial^{2}\chi_{AB}(g\xi^{*},\xi)}{\partial\,z^{2}}\right|_{w,z=0}
\end{equation*}
at the origin of the conjugate phase space, since
$\langle\,\hat{x}\,\rangle_{\widetilde{AB}}=0$.

An identical argument applies to the variance $\langle\,\Delta \hat{p}^{2}\,\rangle_{out}$ which is
given by
\begin{equation} \label{eq:DeltaPOut}
\langle\,\Delta \hat{p}^{2}\,\rangle_{out}\,=\langle\,\Delta
\hat{p}^{2}\,\rangle_{in}\,+\,\langle\,\Delta \hat{p}^{2}\,\rangle_{\widetilde{AB}}
\end{equation}
where, again, we have defined $\langle\,\Delta
\hat{p}^{2}\,\rangle_{\widetilde{AB}}\equiv\langle\,\hat{p}^{2}\,\rangle_{\widetilde{AB}}\,-\,\langle\,\hat{p}\,\rangle^{2}_{\widetilde{AB}}$.
For minimum distortion in $\langle\,\Delta \hat{p}^{2}\,\rangle_{out}$
it is enough to minimize the second derivative
\begin{equation*}
    \langle\,\hat{p}^{2}\,\,\rangle_{\widetilde{AB}}=\left.\frac{\partial^{2}\chi_{AB}(g\xi^{*},\xi)}{\partial\,w^{2}}\right|_{w,z=0}
\end{equation*}
at the origin of the conjugate phase space.

Finally, taking into account~(\ref{eq:XAvgsqout}) and 
\begin{equation}\label{eq:XPAvgsqout}
    \langle\hat{x}\hat{p}\rangle_{out}\,=\,g^{2}\,\langle\,\hat{x}\hat{p}\,\rangle_{in}\,+\,\langle\,\hat{x}\hat{p}\,\rangle_{\widetilde{AB}}\,+\,g\,\langle\,\hat{x}\,\rangle_{in}\,\langle\,\hat{p}\,\rangle_{\widetilde{AB}}+\,g\,\langle\,\hat{p}\,\rangle_{in}\,\langle\,\hat{x}\,\rangle_{\widetilde{AB}}\ ,
\end{equation}
the average for $\mathrm{Cov}(\hat{x},\hat{p})$  is
\begin{equation}
    \mathrm{Cov}(\hat{x},\hat{p})_{out}\,=\,g^{2}\,\mathrm{Cov}(\hat{x},\hat{p})_{in}\,+\mathrm{Cov}(\hat{x},\hat{p})_{\widetilde{AB}}\,
    \label{eq:VarXPOut}
\end{equation}
where we have defined
 $\mathrm{Cov}(\hat{x}
,\hat{p})_{\widetilde{AB}}\,\equiv\,\langle\,\hat{x}\hat{p}\,\rangle_{\widetilde{AB}}\,-\,\langle\,\hat{x}\,\rangle_{\widetilde{AB}}\langle\,\hat{p}\,\rangle_{\widetilde{AB}}$.
For the resources studied is easy to see that $ \langle\,\hat{x}\,\hat{p}\,\rangle_{\widetilde{AB}}=0$ and in fact $\mathrm{Cov}(\hat{x}
,\hat{p})_{\widetilde{AB}}=0$. There is no distortion of $\mathrm{Cov}(\hat{x},\hat{p})_{out}$ produced by a two mode squeezed resource.

\subsubsection{Expectation value of the two-photon correlation function $g^{2}(0)$}

The correlation between the arrival of two successive photons at a photodetector for a given time
interval is proportional to the media of the product of intensities of the electric field at two
different times~\cite{QOptWalls}. For zero-time delay, this is equal to
\begin{equation}\label{eq:g20}
    \mathit{g}^{2}(0)=\frac{\langle\,\hat{a}^{\dagger\;2}\hat{a}^{2}\rangle}{\langle\,\hat{a}^{\dagger}\hat{a}\rangle^{2}}=1+\frac{\mathrm{Var}(\hat{n})-\langle\hat{n}\rangle}{\langle\hat{n}\rangle^{2}}
\end{equation}
with
$\mathrm{Var}(n)=\langle\,(\hat{a}^{\dagger}\hat{a})^{2}\rangle\,-\,\langle\,\hat{a}^{\dagger}\hat{a}\rangle^{2}$
being the photon number  variance. For classical fields,
$g^{2}(0)\geq\,1$.  $g^{2}(0)$, is a measure of the
non-classicality of the photon statistics, fields with non classical photon statistics  and
$g^{2}(0)<1$. It is worthwhile to see, then, how $g^{2}(0)$ changes in \textbf{CV}
teleportation.

The expectation value $\langle\,\hat{a}^{\dagger\;2}\hat{a}^{2}\rangle$ can be written using
equation~(\ref{eq:CharAvgCreAnnOut1}), for a normally-ordered output characteristic function:
\begin{eqnarray}
   &\langle\,\hat{a}^{\dagger\;2}\hat{a}^{2}\rangle_{out}^{(1)}\,=\,g^{4}\langle\hat{a}^{\dagger\;2}\hat{a}^{2}\rangle_{in}^{(1)}+\langle\hat{a}^{\dagger\;2}\hat{a}^{2}\rangle_{\widetilde{AB}}^{(1)}\nonumber\\
   &+2g\left(\langle\hat{a}^{\dagger}\rangle_{in}\langle\hat{a}^{\dagger}\hat{a}^{2}\rangle_{\widetilde{AB}}^{(1)}+\langle\hat{a}\rangle_{in}\langle\hat{a}^{\dagger\;2}\hat{a}\rangle_{\widetilde{AB}}^{(1)}
   \right)\nonumber\\
   &+g^{2}\left(\langle\hat{a}^{\dagger\;2}\rangle_{in}\langle\hat{a}^{2}\rangle_{\widetilde{AB}}+\langle\hat{a}^{\dagger\;2}\rangle_{\widetilde{AB}}\langle\hat{a}^{2}\rangle_{\widetilde{in}}
   \right)\nonumber\\
   &+2g^{3}\left(\langle\hat{a}^{\dagger}\rangle_{\widetilde{AB}}\langle\hat{a}^{\dagger}\hat{a}^{2}\rangle_{in}^{(1)}+\langle\hat{a}\rangle_{\widetilde{AB}}\langle\hat{a}^{\dagger\;2}\hat{a}\rangle_{in}^{(1)} \right) \ .
\end{eqnarray}

The expectation value for $\langle\hat{a}^{\dagger}\hat{a}\rangle$ is given in
equation~(\ref{eq:NOutAvg}). Note that the first order averages
$\langle\hat{a}\rangle_{\widetilde{AB}}$,$\langle\hat{a}^{\dagger}\rangle_{\widetilde{AB}}$
corresponding to the two-mode squeezed resource transfer function (see
equation~(\ref{eq:1stordavtransfereqzero})) are equal to zero. In the following section we show
that the third order averages $\langle\hat{a}^{\dagger}\hat{a}^{2}\rangle_{\widetilde{AB}}^{(1)}$
$\langle\hat{a}^{\dagger\;2}\hat{a}\rangle_{\widetilde{AB}}^{(1)}$ corresponding to the two-mode
squeezed resource transfer function are also equal to zero.

\subsubsection{Expectation value of the central moments of third and fourth order}

To study the distortion caused by teleportation in expectation values, for non-Gaussian input
states, it is necessary to study the expectation values of operators which are third or higher
order in products of momentum and position.

The third order central moment of position for
the teleportation output is given by
\begin{equation}\label{eq:3CentralDef}
\langle\mu_{\hat{x}}^{(3)}\rangle_{out}\,\equiv\,\langle(\hat{x}-\langle\hat{x}\rangle_{out})^{3}\rangle_{out}\,=\,\langle\hat{x}^{3}\rangle_{out}-
\,3\,\langle\hat{x}\rangle_{out}\,\langle\hat{x}^{2}\rangle_{out}+2\langle\hat{x}\rangle_{out}^{3} \ .
\end{equation}

The averages $\langle\hat{x}\rangle_{out}$ and $\langle\hat{x}^{2}\rangle_{out}$ have been worked
out in equations~(\ref{eq:XAvgsqout}) and~(\ref{eq:XXAvgsqout}). From
equation~(\ref{eq:CharAvgXPOut1}) we obtain, for the third order raw moment
\begin{equation}\label{eq:XXXAvgsqout}
 \langle\hat{x}^{3}\rangle_{out}\,=\,g^{3}\langle\hat{x}^{3}\rangle_{in}+3g^{2}\langle\hat{x}^{2}\rangle_{in}\langle\hat{x}\rangle_{\widetilde{AB}}
 \,+\,3g\langle\hat{x}\rangle_{in}\langle\hat{x}^{2}\rangle_{\widetilde{AB}}\,+\,\langle\hat{x}^{3}\rangle_{\widetilde{AB}} \ .
\end{equation}

Using the above results, the third order central moment in equation~(\ref{eq:3CentralDef}) is given by
\begin{eqnarray}
    \langle\mu_{\hat{x}}^{(3)}\rangle_{out}\,&=\,\langle\hat{x}^{3}\rangle_{\widetilde{AB}}+2\langle\hat{x}^{3}\rangle_{\widetilde{AB}}
    -3\langle\hat{x}\rangle_{\widetilde{AB}}\langle\hat{x}^{2}\rangle_{\widetilde{AB}}\nonumber\\
    &+g^{3}\left(\,\langle\hat{x}^{3}\rangle_{in}+2\langle\hat{x}^{3}\rangle_{in}
    -3\langle\hat{x}\rangle_{in}\langle\hat{x}^{2}\rangle_{in}\right)\nonumber \\
    &\,=\,g^{3}\langle\mu_{\hat{x}}^{(3)}\rangle_{in}\,+\,\langle\mu_{\hat{x}}^{(3)}\rangle_{\widetilde{AB}} \label{eq:3CentralDef2}
\end{eqnarray}
and is an additive quantity of the input characteristic function and the transfer
function. This is to be expected, since for any probability distribution up to third order, all the
central moments are identical to cumulants~\cite{Jaynes}. An identical argument to the one outlined
above can be done to obtain, for the third order central moment for $\hat{p}$
\begin{equation}\label{eq:3CentralDef2p}
      \langle\mu_{\hat{p}}^{(3)}\rangle_{out}\,=\,g^{3}\langle\mu_{\hat{p}}^{(3)}\rangle_{in}\,+\,\langle\mu_{\hat{p}}^{(3)}\rangle_{\widetilde{AB}} \ .
\end{equation}

The transfer function for a two-mode squeezed resource, as outlined in
section~(\ref{Averout:transfertwosq}), is of the form $\chi_{AB}(|\xi|^{2}\,e^{-2r})$. This
function has the third order "raw" moment
\begin{eqnarray}
    \langle
\hat{x}^{3}\rangle_{\widetilde{AB}}\,&=&\,\frac{1}{\imath^{3}}\,\left.\left(12\,z\,e^{-4r}\frac{\partial^{2}\chi_{AB}(\gamma)}{\partial\,\gamma^{2}}\,+\,8 z^{3}e^{-6r}\frac{\partial^{3}\chi_{AB}(\gamma)}{\partial\,\gamma^{3}} \right)\right|_{w,z=0}\,  , \label{eq:3ordtransfunc}\\
\gamma&\equiv&|\xi|^{2}\,e^{-2r}\;=\;\xi^{*}\,\xi\,e^{-2r}\;=\;(w^{2}+z^{2})\,e^{-2r}    \nonumber
\end{eqnarray}
which is easily shown to vanish. Identical reasoning shows that $\langle
\hat{p}^{3}\rangle_{\widetilde{AB}}=0$. Inspection of equation~(\ref{eq:3CentralDef}) shows that
for a transfer function with $\langle \hat{p}\rangle_{\widetilde{AB}}=\langle
\hat{x}\rangle_{\widetilde{AB}}=0$, the third order central moments
$\langle\mu_{\hat{x}}^{(3)}\rangle_{\widetilde{AB}}=\langle\mu_{\hat{p}}^{(3)}\rangle_{\widetilde{AB}}=0$.
Therefore, the third order central moments of input states are not changed by teleportation using a
two-mode squeezed resource.

Furthermore, other third order "raw" moments, such as $\langle
\hat{x}^{2}\hat{p}\rangle_{\widetilde{AB}}$ and $\langle
\hat{x}\hat{p}^{2}\rangle_{\widetilde{AB}}$ are also zero. For example
\begin{equation}\label{eq:3rawx2p}
\langle\hat{x}^{2}\hat{p}\rangle_{\widetilde{AB}}\,=\,\frac{1}{\imath^{3}}\,\left.\frac{\partial}{\partial
z^{2}}\frac{\partial\,\chi_{AB}(\gamma)}{\partial \gamma}\;2\,w\,e^{-2r}\right|_{w,z=0}=0
\end{equation}
where $\gamma$ is defined in equation~(\ref{eq:3ordtransfunc}). Analogous calculations show that
$\langle\hat{x}\hat{p}^{2}\rangle_{\widetilde{AB}}$ vanishes.

Finally, third order moments of the form
$\langle\hat{a}^{\dagger\,n}\hat{a}^{m}\rangle_{\widetilde{AB}}$ with $n+m=3$ can be written as
linear combinations of third order moments of $\hat{x}$ and $\hat{p}$ and therefore  vanish,
regardless of operator ordering considerations. Observable quantities such as
$\mathit{g}^{2}(0)_{out}$ (see equation~(\ref{eq:g20})), which are of fourth order in
$\hat{a}^{\dagger}$ and $\hat{a}$, depend  only on the second and fourth order moments associated
with the transfer function and the input state.

To further study distortion through teleportation fourth order central moments must be analyzed.
The fourth-order central moments are not identical to cumulants~\cite{Jaynes}. and hence the fourth
order central moment of the teleportation output is not equal to the sum of the fourth order
central moments of input and resource. From
\begin{eqnarray}\label{eq:4CentralXDef}
\langle\mu_{\hat{x}}^{(4)}\rangle_{out}\,&\equiv&\,\langle(\hat{x}-\langle\hat{x}\rangle_{out})^{4}\rangle_{out}\,\ \nonumber\\
&=&\,\langle\hat{x}^{4}\rangle_{out}-
\,4\,\langle\hat{x}\rangle_{out}\,\langle\hat{x}^{3}\rangle_{out}\,+6\langle\hat{x^{2}}\rangle_{out}\langle\hat{x}\rangle_{out}^{2}
-3\langle\hat{x}\rangle_{out}^{4} \ ,
\end{eqnarray}
\begin{eqnarray}\label{eq:XXXXAvgsqout}
    \langle\hat{x}^{4}\rangle_{out}&=&g^{4}\langle\hat{x}^{4}\rangle_{in}
    \,+ \,4\,g^{3}\langle\hat{x}^{3}\rangle_{in}\langle\hat{x}\rangle_{\widetilde{AB}}
    \,+\,6\,g^{2}\langle\hat{x}^{2}\rangle_{in}\langle\hat{x}^{2}\rangle_{\widetilde{AB}}
    \,\nonumber\\
&+&\,4\,g\langle\hat{x}\rangle_{in}\langle\hat{x}^{3}\rangle_{\widetilde{AB}}
    \,+\,\langle\hat{x}^{4}\rangle_{\widetilde{AB}} \ ,
\end{eqnarray}
it is shown that
\begin{equation}\label{eq:4CentralDefXout}
\langle\mu_{\hat{x}}^{(4)}\rangle_{out}\,=\,\langle\mu_{\hat{x}}^{(4)}\rangle_{out}\,+\,\langle\mu_{\hat{x}}^{(4)}\rangle_{\widetilde{AB}}
\,+\, 6 g^{2} \langle \Delta \hat{x}^{2}\,\rangle_{in}\,\langle\Delta
\hat{x}^{2}\,\rangle_{\widetilde{AB}}
\end{equation}
where $\langle\Delta \hat{x}^{2}\,\rangle_{in}$ and $\langle\Delta \hat{x}^{2}\,\rangle_{AB}$ are
second order central moments. The fourth order central moment for $\hat{p}$ is calculated in an
analogous manner;
\begin{equation}\label{eq:4CentralPDefout}
\langle\mu_{\hat{p}}^{(4)}\rangle_{out}\,=\,\langle\mu_{\hat{p}}^{(4)}\rangle_{out}\,+\,\langle\mu_{\hat{p}}^{(4)}\rangle_{\widetilde{AB}}
\,+\, 6 g^{2} \langle \Delta \hat{p}^{2}\,\rangle_{in}\,\langle\Delta
\hat{p}^{2}\,\rangle_{\widetilde{AB}} \ .
\end{equation}

The distortion caused by teleportation on the fourth order central moments of the output is a sum
of the fourth order central moment of the transfer function and a product of the second order
central moments of the input and transfer function. Thus it is not additive in like order central
moments.

The fourth order cumulant of the output state defined in terms of fourth and second order central
moments is  additive in the corresponding cumulants of the input and transfer
function~\cite{Jaynes}.  We have,
\begin{equation}\label{eq:Cum4XDef}
\langle\kappa^{(4)}_{\hat{x}}\rangle\,\equiv\,\langle\mu_{\hat{x}}^{(4)}\rangle\:-\:3\,(\langle
\Delta \hat{x}^{2}\,\rangle)^{2}
\end{equation}
and
\begin{equation}\label{eq:Cum4PDef}
\langle\kappa^{(4)}_{\hat{p}}\rangle\,\equiv\,\langle\mu_{\hat{p}}^{(4)}\rangle\:-\:3\,(\langle
\Delta \hat{p}^{2}\,\rangle)^{2} \ .
\end{equation}

As already said
\begin{eqnarray}
\langle\kappa^{(4)}_{\hat{x}}\rangle_{out}\,&=&\,\langle\kappa^{(4)}_{\hat{x}}\rangle_{\widetilde{AB}}\,+\,g^{4}\langle\kappa^{(4)}_{\hat{x}}\rangle_{in}\label{eq:Cum4XOut}\\
\langle\kappa^{(4)}_{\hat{p}}\rangle_{out}\,&=&\,\langle\kappa^{(4)}_{\hat{p}}\rangle_{\widetilde{AB}}\,+\,g^{4}\langle\kappa^{(4)}_{\hat{p}}\rangle_{in} \ . 
\label{eq:Cum4POut} 
\end{eqnarray}

For $g=1$, the distortion caused by teleportation on the $\hat{x}$ ($\hat{p}$) fourth order
cumulant of the teleported state is equal to the fourth order cumulant of the transfer function. As
with the case of lower order cumulants, the distortion depends on the resource properties only.
Furthermore, the transfer function for the two-mode squeezed state described in
subsection~(\ref{Averout:transfertwosq}) is isotropic in the conjugate phase space with coordinate
$\xi=w\,+\,\imath\,z$ and thus
$\langle\kappa^{(4)}_{\hat{x}}\rangle_{\widetilde{AB}}=\langle\kappa^{(4)}_{\hat{p}}\rangle_{\widetilde{AB}}$.
Optimization of teleportation distortion of fourth order cumulants is performed through
minimization of $\langle\kappa^{(4)}_{\hat{x}}\rangle_{\widetilde{AB}}$ alone.

\section{Expectation Values Differences and Optimization of Transfer Functions}\label{OptAvEx}

We consider now  teleportation  of chosen input states with the 
teleportation resource given by Squeezed Bell-like states defined by,
\begin{align}
\chi_{sbl}(\xi_{A} ;\: \xi_{B}) = & \,
e^{-1/2\,\left(|\,\xi'_{A}\,|^{2}\,+\,|\,\xi'_{B}\,|^{2}\right)}\: \{ \Delta^{2} \:+\: 2
\Delta\,\sqrt{1-\Delta^{2}}\, \mathrm{Re}[e^{\,i\, \theta} \xi'_{A}\, \xi'_{B}]
\notag \\
 & \,+ \, (1-\Delta^{2})\,(1\,-\,|\,\xi'_{A}\,|^{2})\,(1\,-\,|\,\xi'_{B}\,|^{2}) \}\ \ , 
\label{eq:CharSqueBell}
\end{align}
with $\xi'_{A/B} =\: \cosh(r)\, \xi_{A/B}\, - \sinh(r)\, \xi_{B/A}^{*}$ and fixed squeezing. They
depend on a free parameter $\Delta$ which can be used to optimize the teleportation output. This
was done in ~\cite{TelepNonGauss} to optimize fidelity. Here we consider other, more general
properties of the output and compare the results.

We consider the following input states: A Fock state with photon number $1$, a
coherent state with a real displacement $\beta$ and a squeezed vacuum state with squeezing $s$.
Their characteristic functions are, respectively
\begin{eqnarray}
\chi_{f1}(\xi_{in}) \,&=&\, e^{\,-\,\frac{1}{2}|\,\xi_{in}\,|^{2}}\:(1\,-\,|\,\xi_{in}\,|^{2})
\label{eq:CharSqFockIn}\\
\chi_{coh,\beta}(\xi_{in}) \,&=&\, e^{-\frac{1}{2}|\,\xi_{in}\,|^{2}\,+\,2\,i\,
\mathrm{Im}[\xi_{in}]\beta} \label{eq:CharCohIn}\\
\chi_{sqc}(\xi_{in}) \,&=&\, e^{-\frac{1}{2}|\,\xi'_{in}\,|^{2}}
\label{eq:CharSqVacIn}\\
\xi_{in}'\,&=&\,\xi_{in} \cosh(s) \,+\, \xi_{in}^{*}\sinh(s) \ .  \nonumber\\
\end{eqnarray}

The  value of $\Delta$ for optimal fidelity for a Fock state and a coherent state (and vacuum
state) were computed in~\cite{TelepNonGauss} for  $\theta=0$ and $g=1$,
\begin{eqnarray}
\Delta_{f1}^{opt} \,&=&\,\cos\left( \frac{1}{2}\,\arctan\left(
\frac{e^{-2\,r}(1-e^{2\,r}+e^{4\,r}+3e^{6\,r})}{3(e^{2\,r}-1)^{2}} \right) \right)
\label{eq:DeltaSqbOptFock}\\
\Delta_{coh}^{opt}  \,&=&\,\cos\left(\frac{1}{2}\,\arctan(1\,+\,e^{-2\,r}\,)\right) \ . 
\label{eq:DeltaSqbOptCoh}
\end{eqnarray}
The squeezed vacuum optimal fidelity parameter $\Delta_{sqc}^{opt}$ was also found
in Reference ~\cite{TelepNonGauss}.

\subsection{Teleportation of Covariance Matrix elements} \label{OptAvEx:CovMat}

Consider the difference between the average $\langle\,\Delta
\hat{x}^{2}\,\rangle_{out}$ and $\langle\,\Delta \hat{x}^{2}\,\rangle_{in}$;
\begin{equation}\label{eq:DifVarX}
    \mathrm{D}_{\Delta
\hat{x}^{2}}\,\equiv\,\left|\langle\,\Delta
\hat{x}^{2}\,\rangle_{out}-\langle\,\Delta \hat{x}^{2}\,\rangle_{in}\right| \ .
\end{equation}

From equation~(\ref{eq:DeltaXOut}), we have
\begin{equation}\label{eq:DifVarX2}
    \mathrm{D}_{\Delta
\hat{x}^{2}}\,=\,\left|\langle\,\Delta \hat{x}^{2}\,\rangle_{\widetilde{AB}}\right| \ .
\end{equation}

\begin{figure}[tb]
\centering
\includegraphics[width= 0.6 \linewidth]{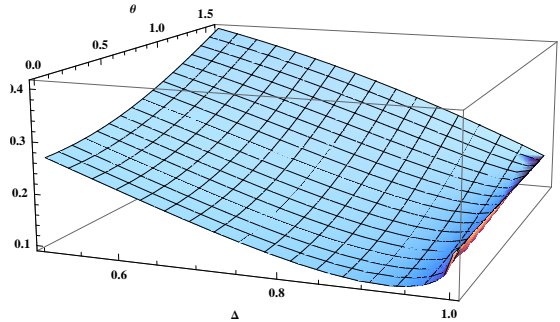}
\caption{$\langle\hat{x}^{2}\rangle_{\widetilde{AB}}$ as a function of  $\Delta$ and
$\theta$ for $r=1.25$} \label{fig:XXSqb}
\end{figure}

This does not depend on the input. We note that for 
Squeezed Bell-like states the transfer function depends actually on $e^{-2r}|\xi|^{2}=e^{-2r}(w^{2}+z^{2})$. We have $ \langle \hat{x}^{2}\rangle_{\widetilde{AB}}\,=\,\langle \hat{p}^{2}\rangle_{\widetilde{AB}}$ with
\begin{equation}\label{eq:XXSqb}
 \langle \hat{x}^{2}\rangle_{\widetilde{AB}}\,=\,\left.\frac{\partial^{2}\chi_{AB}(g\xi^{*},\xi)}{\partial\,z^{2}}\right|_{w,z=0}\,=\,e^{-2 r} \left(6-4 \Delta ^2-4 \Delta  \sqrt{1-\Delta ^2} \cos(\theta)\right) \ .
\end{equation}
 Furthermore,  $\langle \hat{x}\rangle_{\widetilde{AB}}$,$\langle
\hat{p}\rangle_{\widetilde{AB}}$ as well as $\mathrm{Cov}(\hat{x}\hat{p})_{\widetilde{AB}}$ vanish. We need only to minimize  $\langle \hat{x}^{2}\rangle_{\widetilde{AB}}$ which is displayed for a
fixed value of $r$  in figure~(\ref{fig:XXSqb})
This function, for $\theta=0$ has a minimum at
\begin{equation}\label{eq:MagNumOptCov}
    \Delta_{(2)}^{opt}\,=\,\frac{\sqrt{2+\sqrt{2}}}{2}\,=\,0.92388 \ ,
\end{equation}
which is independent of the two-mode squeezing $r$ of the resource, although the value of $\langle
\hat{x}^{2}\rangle_{\widetilde{AB}}$ for the Squeezed Bell-like resource goes to $0$ as
$r\rightarrow\infty$. We note  also that the optimum fidelity resource parameters for the
coherent, Fock and squeezed vacuum inputs converge towards the same value when
$r\rightarrow\infty$;
\begin{eqnarray}
\lim_{r\rightarrow\infty}\,\Delta_{f1}^{opt} \,=\,\lim_{r\rightarrow\infty}\,\Delta_{coh}^{opt}
\,=\,\lim_{r\rightarrow\infty}\Delta_{sqc}^{opt}\,=\Delta_{(2)}^{opt}\label{eq:MagNumOptFidConv}\ \ .
\end{eqnarray}

For large squeezing of the resource the distortion in teleportation of the second order momenta
dominates the fidelity behavior.

\subsection{Teleportation  of Squeezing}\label{OptAvEx:Squeez}

The output state of teleportation has a squeezing equal to
\begin{equation}\label{eq:EquivSqueezTeleport}
    \mathcal{S}_{out}=\frac{\langle\,\Delta\,\hat{x}^{2}\,\rangle_{out}}{\langle\,\Delta\,\hat{p}^{2}\,\rangle_{out}}\,
    =\,\frac{\langle\,\Delta\,\hat{x}^{2}\,\rangle_{in}+\langle\,\Delta\,\hat{x}^{2}\,\rangle_{\widetilde{AB}}}{\langle\,\Delta\,\hat{p}^{2}\,\rangle_{in}
    +\langle\,\Delta\,\hat{p}^{2}\,\rangle_{\widetilde{AB}}}\ \ .
\end{equation}
For the two-mode squeezed resources we are considering
$\langle\,\Delta\,\hat{x}^{2}\,\rangle_{\widetilde{AB}}=\langle\,\Delta\,\hat{p}^{2}\,\rangle_{\widetilde{AB}}$.
Input states that  initially are not squeezed remain unsqueezed and
initially squeezed input states will become less
squeezed after teleportation.

\begin{figure}[tb]
\centering
\includegraphics[width=0.6 \linewidth]{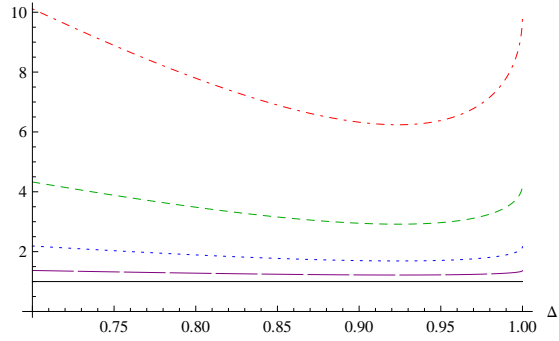}
\caption{The quotient of Squeezing $\mathcal{S}_{out}/\mathcal{S}_{in}$ for a squeezed vacuum state
with $s=0$ (solid line), $s=0.5$ (long dashed line), $s=1.0$ (dotted line), $s=1.5$ (short dashed
line) and $s=2.0$ (dotted-dashed line) as a function of the $\Delta$ parameter of the Squeezed
Bell-like resource.The two-mode squeezing of the resource has been fixed at $r=1.25$}
\label{fig:DifEquSqsqc}
\end{figure}

In figure~(\ref{fig:DifEquSqsqc}) the quotient $\mathcal{S}_{out}/\mathcal{S}_{in}$ between the
squeezing of the input and output states of a squeezed vacuum state is shown, for a Squeezed
Bell-like resource with  squeezing $r=1.25$, and for several values the  squeezing
parameter $s$ of the input state. For the inputs displayed $\mathcal{S}_{in}=e^{-4s}$
and $1\geq\mathcal{S}_{out}\geq\mathcal{S}_{in}$. Note that the quotient is greater for higher
values of $s$.

The Squeezed Bell-like resource minimizing $\mathcal{S}_{out}/\mathcal{S}_{in}$ is the one that
minimizes the distortion in the output state covariance matrix elements
$\langle\,\Delta\,\hat{x}^{2}\,\rangle_{out}$ and $\langle\,\Delta\,\hat{p}^{2}\,\rangle_{out}$ ,
and has $\Delta=\Delta_{(2)}^{opt}$.

\subsection{Teleportation of Photon Number}\label{OptAvEx:NumPhot}

Consider now the photon number average given by equation~(\ref{eq:NOutAvg}). For a two-mode
squeezed resource,  "averages" of the transfer function such as
$\langle\,\hat{a}\,\rangle_{\widetilde{AB}}=\langle\,\hat{a}^{\dagger}\,\rangle_{\widetilde{AB}}$,
vanish as discussed in section~(\ref{Averout:transfertwosq}). The teleportation distortion of
photon number average depends only on the resource, regardless of input. The optimization of photon
number transmission  is carried out by minimizing the photon number "average" of the transfer
function.  This is done by minimizing the second derivative
\begin{equation*}
    \langle\,\hat{a}^{\dagger}\,\hat{a}\,\rangle_{\widetilde{AB}}=\left.\frac{\partial^{2}\chi^{(1)}_{AB}(\xi^{*},\xi)}{\partial\,\xi\;\partial\,\xi}\right|_{\xi,\xi^{*}=0}
\end{equation*}
at the origin of the conjugate phase space. The photon number "average" for the transfer function
involves a second derivative over $\xi=w\,+\,\imath\,z$ and $\xi^{*}$. Thus, it is a function of
identical behavior to that of equation~(\ref{eq:XXSqb}). It is given by
\begin{equation}\label{eq:NSqb}
 \langle \hat{a}^{\dagger}\,\hat{a}\rangle_{\widetilde{AB}}\,=\,-e^{-2 r} \left(-3+e^{2 r}+2 \Delta ^2+2 \Delta  \sqrt{1-\Delta ^2} \cos(\theta)\right)
\end{equation}
and for $\theta=0$, it has a minimum at $\Delta\,=\Delta_{(2)}^{opt}$.

\subsection{Teleportation of Fourth Order Central Moments and Cumulants}

As discussed above, the fourth order central moments for position and momentum of the output state
are not additive in the central moments of input and transfer function. Therefore, the distortion
caused by teleportation depends on the properties of the input state. Let us measure this
distortion as the absolute value of the difference between the central momenta,
\begin{equation}\label{eq:Dif4CentX}
    \mathrm{D}_{\mu_{\hat{x}}^{(4)}}\,\equiv\,\left|\langle\mu_{\hat{x}}^{(4)}\rangle_{out}-\langle\mu_{\hat{x}}^{(4)}\rangle_{in}\right|\,=\,
    \left|\langle\mu_{\hat{x}}^{(4)}\rangle_{\widetilde{AB}}
\,+\, 6 \langle \Delta \hat{x}^{2}\,\rangle_{in}\,\langle\Delta
\hat{x}^{2}\,\rangle_{\widetilde{AB}} \right| \ .
\end{equation}

The optimal values of $\Delta$  in this case depend on the two-mode squeezing $r$ of the Squeezed Bell-like
resource. For the squeezed vacuum input there is a dependence on the squeezing $s$. We have,
\begin{eqnarray}
\Delta_{coh,{(4)}}^{\mu_{\hat{x}}}\,&=&\,\frac{\sqrt{1+\frac{\left(3+e^{2
r}\right)^2}{\sqrt{\left(3+e^{2 r}\right)^2 \left(13+2 e^{2 r} \left(5+e^{2
r}\right)\right)}}}}{\sqrt{2}}\label{eq:DeltaOpt4Xcoh}\\
\Delta_{sqc,{(4)}}^{\mu_{\hat{x}}}\,&=&\,\frac{\sqrt{1+\frac{\left(e^{2 r}+3 e^{2
s}\right)^2}{\sqrt{\left(e^{2 r}+3 e^{2 s}\right)^2 \left(2 e^{4 r}+13 e^{4 s}+10 e^{2
(r+s)}\right)}}}}{\sqrt{2}}\label{eq:DeltaOpt4Xsqc}\\
\Delta_{f1,{(4)}}^{\mu_{\hat{x}}}\,&=&\,\frac{\sqrt{1+\frac{3 \left(1+e^{2
r}\right)^2}{\sqrt{\left(1+e^{2 r}\right)^2 \left(13+30 e^{2 r}+18 e^{4
r}\right)}}}}{\sqrt{2}} \ .
\label{eq:DeltaOpt4Xf1} 
\end{eqnarray}

The squeezed vacuum state is not isotropic in phase-space and has different central moments in
the powers of $\hat{x}$  and $\hat{p}$. Optimization for the momentum $\hat{p}$  distortion of the
fourth order central moment yields
\begin{equation}\label{eq:DeltaOpt4Psqc}
    \Delta_{sqc,{(4)}}^{\mu_{\hat{p}}}\,=\,\frac{\sqrt{1+\frac{\left(3+e^{2 (r+s)}\right)^2}{\sqrt{\left(3+e^{2 (r+s)}\right)^2 \left(13+2 e^{2 (r+s)} \left(5+e^{2 (r+s)}\right)\right)}}}}{\sqrt{2}}\ \ .
\end{equation}

For $s=0$ the equations~(\ref{eq:DeltaOpt4Xsqc}) and~(\ref{eq:DeltaOpt4Psqc}) reduce to
equation~(\ref{eq:DeltaOpt4Xcoh}). The limit for $r\rightarrow\infty$  in
equations~(\ref{eq:DeltaOpt4Xcoh}),~(\ref{eq:DeltaOpt4Xsqc}),~(\ref{eq:DeltaOpt4Xf1})
and~(\ref{eq:DeltaOpt4Psqc}) is  $\Delta_{(2)}^{opt}$.

\begin{figure}[tb]
\centering
\includegraphics[width=0.6 \linewidth]{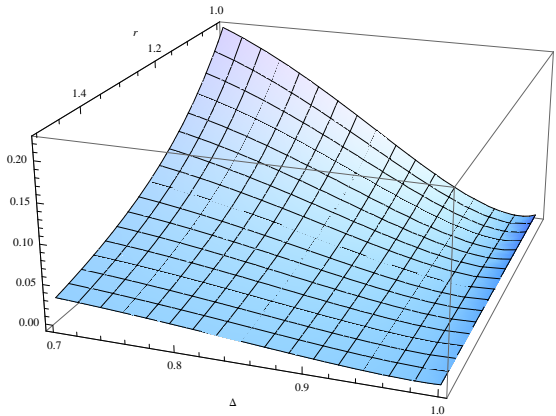}
\caption{The fourth order cumulant $\langle\kappa^{(4)}_{\hat{x}}\rangle_{\widetilde{AB}}$
of the transfer function of a Squeezed Bell-like resource with phase $\theta=0$, as a function of $\Delta$ and $r$, } \label{fig:Cum4Sbl}
\end{figure}

The transfer function of a non-Gaussian resource such as the squeezed Bell-like state has non
vanishing fourth order central moments (and cumulants). Teleportation using this resource changes
the fourth order central moments (and cumulants) of any input. Particularly, for a Gaussian input
the output will be non-Gaussian. On the other hand for $g=1$, the teleportation distortion in
fourth order ($\hat{x}$ and $\hat{p}$) cumulants  is determined by
$\langle\kappa^{(4)}_{\hat{x}}\rangle_{\widetilde{AB}}$. For the squeezed Bell-like resource with
phase $\theta=0$, this is given by
\begin{equation}\label{eq:Cum4Sbl}
\langle\kappa^{(4)}_{\hat{x}}\rangle_{\widetilde{AB}}\,=\,24 e^{-4 r} \left(-1+\Delta ^2\right)
\left(1-4 \Delta  \sqrt{1-\Delta ^2}\right)
\end{equation}
and is displayed in figure~(\ref{fig:Cum4Sbl}). Numerical optimization with respect to the $\Delta$
parameter at fixed $r$ yields an optimal value of
\begin{equation}\label{eq:DeltaOpt4Cum}
    \Delta_{(4)}^{opt}\approx 0.985294 \ .
\end{equation}

The fourth derivatives with respect to $w$ and $z$ of the transfer function for a two-mode squeezed
state at $w,z=0$ go to zero for large $r$ faster than the second derivatives. This explains the
convergence of optimal fidelity resources  when $r\rightarrow\infty$ to the optimal resources for
teleportation of second order moments.

Cumulants of higher order than fourth go to zero at a faster rate than the fourth order cumulant.
Therefore, for large $r$ and approximate \textbf{EPR} resources the distortion of the teleported
state is determined by distortion of second order central moments, and resources optimal for second order
moments optimize fidelity too.

%%%%%%%%%%%%%%%%%%%%%%%%%%%%Finis%%%%%%%%%%%%%%%%%%%%%%%%%%%%%%%%%%%%%%%%%%%%%
\section{Teleportation of Photon Statistics}\label{Dfunc}

Let us focus now in the differences in the statistics of input and output measurements for a given
observable. A readily measurable quantity in \textbf{CV} is the photon number observable,
associated with the operator $\hat{n}=\hat{a}^{\dagger}\hat{a}$, and its eigenstates  $\{|n\,\rangle\;n=0,1,2,\ldots,\infty\}$. 

\subsection{Deviations in photon statistics}\label{Dfunc:Def}
Consider the functional
 \begin{equation}\label{eq:DFun0}
    \mathcal{D}_{N}\equiv\,\left(\sum_{n=0}^{N}\;\left(\,P_{n}^{(out)}-\,P_{n}^{(in)}\right)^{2}\right)^{\frac{1}{2}}
\end{equation} where the deviations are summed up to an arbitrary finite photon number $N$, which
makes $\mathcal{D}_{N}$ numerically computable. For a quantum state described by a density operator
$\hat{\rho}$ and a Wigner characteristic function $\chi(\xi)$ the photon number probability is
given by
\begin{equation}\label{eq:ProbPhotChar}
    P_{n}\,=\,\mathrm{Tr}\left(\hat{\rho}\,\hat{\rho}_{n}\right)\,=\,\frac{1}{\pi}\int\,d^{2}\xi\,\chi(\xi)\,\chi_{n}(-\xi)
\end{equation} where $\hat{\rho}_{n}$ and $\chi_{n}(\xi)$ are respectively, the density operator
and the characteristic function of the Fock state $|n\,\rangle$.

For a quantum state with  finite energy (and square-integrable Wigner and Wigner characteristic
functions)  $P_{N}\rightarrow\,0$ when $N\rightarrow\,\infty$. Therefore, the sum $\mathcal{D}_{N}$
converges for large $N$. An upper bound for $\mathcal{D}_{N}$ is obtained when input and output
are two different Fock states. Then $\mathcal{D}_N=\sqrt{2}$.

Identical photon statistics  corresponding to $\mathcal{D}_{N}=0$ does not imply identical input and
output states, since the phases of the complex amplitudes  can be different. For example, two
coherent states $|\beta\,\rangle$ and $|\beta\,e^{i\,\phi}\,\rangle$  have identical photon
statistics
\begin{equation*}
P_{n}=e^{-|\beta|^{2}}\;\frac{|\beta|^{2n}}{n!}\ \ ,
\end{equation*} though they
are different states and can be nearly orthogonal for large $|\beta|$ and a considerable phase
difference $\phi$. A simpler example is constructed with  the following superpositions of the $\left\{|0\rangle\,,\,|1\rangle\right\}$
\begin{eqnarray*}
|\Psi_{+}\,\rangle=\frac{1}{\sqrt{2}}\left(|0\,\rangle\;+\;|1\,\rangle\right)\ \ ,\ \
|\Psi_{-}\,\rangle= \frac{1}{\sqrt{2}}\left(|0\,\rangle\;-\;|1\,\rangle\right)
\end{eqnarray*}
which have  $P_{0}=P_{1}=1/2$ and $\mathcal{D}_{N}=0$.

\subsection{Increment and convergence with increasing $N$}\label{Dfunc:Conv}

The increment in $\mathcal{D}_{N}$ associated with  an additional $N+1$-photon term to the sum can
be approximated by differentiation. Let

\begin{eqnarray}
\delta_{N}&\equiv&\,\left(P_{N+1}^{(out)}-P_{N+1}^{(in)}\right)^{2} \label{eq:DeltaDFun} \\
\mathcal{D}_{N+1}&=&\left(\sum_{n=0}^{N}\;\left(\,P_{n}^{(out)}-\,P_{n}^{(in)}\right)^{2}+\left(\,P_{N+1}^{(out)}-\,P_{N+1}^{(in)}\right)^{2}\right)^{\frac{1}{2}}\nonumber \\
&=&\left(\mathcal{D}_{N}^{2}+\delta_{N}\right)^{\frac{1}{2}}=\mathcal{D}_{N}\,\left(1+\frac{\delta_{N}}{\mathcal{D}_{N}^{2}}\right)^{\frac{1}{2}} \ .
\label{eq:DFunNp1}
\end{eqnarray}

This can be rewritten as a Taylor series of $\mathcal{D}_{N+1}$ with respect
to $\delta_{N}$;
\begin{equation}\label{eq:DFunNp1Taylor}
\mathcal{D}_{N+1}\,=\,\mathcal{D}_{N}\,+\,\frac{1}{2\,\mathcal{D}_{N}}\,\delta_{N}\,-\,\frac{1}{8\,\mathcal{D}_{N}^{3}}\,\delta_{N}^{2}\,
+\,\frac{3}{48\,\mathcal{D}_{N}^{5}}\,\delta_{N}^{3}\,-\,\frac{15}{384\,\mathcal{D}_{N}^{7}}\,\delta_{N}^{4} \ .
\end{equation}

For large enough $N$, $\delta_{N}\ll\mathcal{D}_{N}^{2}$, and quadratic (and higher order) terms in
equation~(\ref{eq:DFunNp1Taylor}) are negligible. Therefore,
\begin{equation}\label{eq:DFunNp1Appr}
\mathcal{D}_{N+1}\,\approx\,\mathcal{D}_{N}+\frac{\delta_{N}}{2\,\mathcal{D}_{N}}\,
=\,\mathcal{D}_{N}+\frac{1}{2\,\mathcal{D}_{N}}\;\left(P_{N+1}^{(out)}-\,P_{N+1}^{(in)}\right)^{2}\ .
\end{equation}
We have a positive, decreasing increment for increasing $N$. Therefore, $\mathcal{D}_{N}$ converges
from below.

\subsection{Fidelity and Frobenius Distance} \label{CompD:Fidelity}

Fidelity between two quantum states, $\hat{\rho}^{in}$ and $\hat{\rho}^{out}$ is given by
\begin{equation}\label{eq:FidelityFock}
    F\left[\hat{\rho}^{in},\hat{\rho}^{out}\right]\,=\,\mathrm{Tr}\left(\hat{\rho}^{in}\hat{\rho}^{out}\right)=\sum_{n}\,\sum_{m}\rho^{in}_{nm} \, \rho^{out}_{mn}
\end{equation}
which is equal to the square of the scalar product of the input and output states, when both states
are pure. It is symmetric to the exchange of density matrices, has an upper bound equal to the
purity of the purest state involved and a lower bound  equal to $0$. Let
$\hat{\rho}^{in}=\hat{\rho}^{out}$, then,
\begin{equation}\label{eq:Fidepurity}
    F\left[\hat{\rho}^{in},\hat{\rho}^{in}\right]=\mathrm{Tr}\left(\hat{\rho}^{in\:2}\right)\,\leq\,1 \ .
\end{equation}

Frobenius distance between the same two states is given by
\begin{eqnarray}\label{eq:Frobenius1}
    \delta_{F}&=&\left(\mathrm{Tr}\left((\hat{\rho}^{out}-\hat{\rho}^{in})^{\dagger}(\hat{\rho}^{out}-\hat{\rho}^{in})\right)\,\right)^{1/2}\,\nonumber \\
    &=&\,\left(\mathrm{Tr}\left(\hat{\rho}^{in\:2}\right)+\mathrm{Tr}\left(\hat{\rho}^{out\:2}\right)\,-\,2\,\mathrm{Tr}\left(\hat{\rho}^{in}\hat{\rho}^{out}\right)\right)^{1/2} \ ,
    \nonumber \\
    &=&\,\left(\mathrm{Tr}\left(\hat{\rho}^{in\:2}\right)+\mathrm{Tr}\left(\hat{\rho}^{out\:2}\right)\,-\,2\,F\left[\hat{\rho}^{in},\hat{\rho}^{out}\right]\right)^{1/2} \ .
\end{eqnarray}

Fidelity can be compared directly to  $\mathcal{D}_{N}$ and with the Frobenius distance between
the two states $\hat{\rho}^{in}$ and $\hat{\rho}^{out}$  when one of the states (for instance
$\hat{\rho}^{in}$)  is a mixture of Fock states, with a diagonal density matrix in the Fock basis.

Let
\begin{equation}
\hat{\rho}^{in}=\,\sum_{k}\,P_{k}^{(in)}\,|k\,\rangle\,\langle\,k\,|\label{eq:FockMixIn}\ .
\end{equation}
For this state, fidelity is given by
\begin{eqnarray}
    F\left[\hat{\rho}^{in},\hat{\rho}^{out}\right]&=&\sum_{n,m}\,\sum_{k}\,P_{k}^{(in)}\,\langle\,n|\,k\,\rangle\,\langle\,k\,|\,m\,\rangle\,\langle\,m\,|\hat{\rho}^{out}|\,n\,\rangle\\
    &=&\sum_{k}P_{k}^{(in)}\rho^{out}_{kk}=\,\sum_{k}P_{k}^{(in)}\,P_{k}^{(out)} \ .
\label{eq:FideFockmix}
\end{eqnarray}

Then, 
\begin{equation}\label{eq:CompFidFockD}
    \mathcal{D}_{N}=\left(\sum_{k}\;(P_{k}^{(out)})^{2}\,+\,(P_{k}^{(in)})^{2}\,-\,2\,F\left[\hat{\rho}^{in},\hat{\rho}^{out}\right]\right)^{1/2}\ .
\end{equation}

Using equation~(\ref{eq:FideFockmix}) Frobenius distance can be rewritten;
\begin{equation}\label{eq:Frobenius2}
    \delta_{F}=
    \left(\mathrm{Tr}\left((\hat{\rho}^{in})^{2}\right)+\mathrm{Tr}\left((\hat{\rho}^{out})^{2}\right)\,-\,2\,F\left[\hat{\rho}^{in},\hat{\rho}^{out}\right]\right)^{1/2} \ .
\end{equation}

This  is equal to  $\mathcal{D}_N$  for the special
case in which both of the states are mixtures of pure Fock states. For inputs which are mixtures of
pure Fock states,  $\mathcal{D}_N$  approximates  Frobenius distance closely and
therefore, $\mathcal{D}_N$ is a good measure of the accuracy of teleportation.

\section{Computation of $\mathcal{D}_N$}\label{DFuncSamples}

In this section we compute  $\mathcal{D}_N$ of equation~(\ref{eq:DFun0}), up to
photon number $24$ for sample input states  and their  output states teleported using a Squeezed
Bell-like resource with varying two-mode squeezing $r$. The photon statistics for the output state
are calculated by numerical integration of the matrix element
\begin{equation}\label{eq:photnumbout}
    P_{n}=\hat{\rho}^{out}_{nn}=\pi^{-1}\,\int\,d^{(2)}\xi\chi_{in}(\xi)\,\chi_{AB}(\xi^{*};\xi)\chi_{n}(-\xi)
\end{equation}
where $\chi_{n}(\xi)$ is the characteristic function for a Fock state with photon number $n$. For
the input states the exact photon number probabilities are used.

$\mathcal{D}_N$ has been obtained to an accuracy of $7$ digits, and found to have a first-order
differential from $24$ to $25$ photons  that is, for all inputs
and for the resource squeezing $r$ reported, at least two orders of magnitude smaller than $\mathcal{D}_N$. For comparison purposes in
the figures, $\mathcal{D}_N$ is displayed alongside
$1-F\left[\hat{\rho}^{in},\hat{\rho}^{out}\right]$ (not to scale).

$\mathcal{D}_N$  has been analyzed for Fock states, with photon number $0$ (vacuum
state) and photon number $1$ (equation~(\ref{eq:CharSqFockIn})); a mixture of the two Fock states
with photon numbers $0$ and $1$ of equal probability $P_{0}^{(in)}=P_{1}^{(in)}=0.5$, a coherent
state (equation~(\ref{eq:CharCohIn})) with displacement  $\beta=2.12928$ and a squeezed vacuum (see
equation~(\ref{eq:CharSqVacIn})) with squeezing equal to $s=1.5$

\subsection{Vacuum Input State}\label{DFuncSamples:Vacuum}

\begin{figure}[tb]
\centering
%%----primera----
\subfloat[$r=0.75$]{
\includegraphics[width=0.35\linewidth]{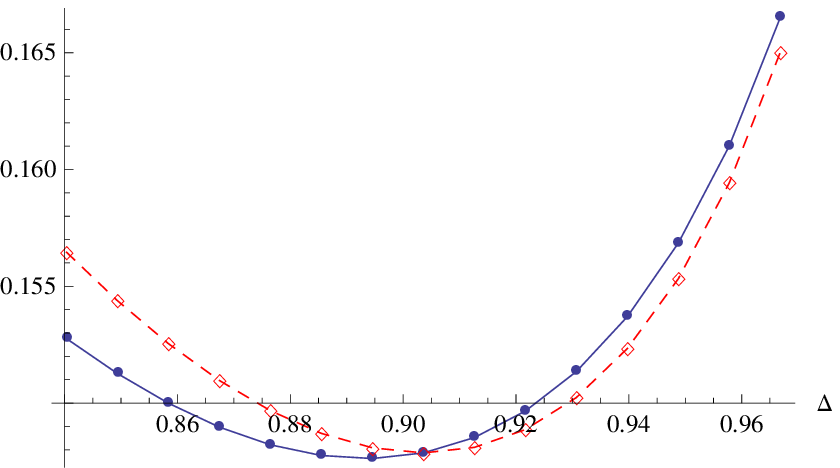}}
\hspace{0.1\linewidth}
%%----start of second subfigure----
\subfloat[$r=1.00$]{
\includegraphics[width=0.35\linewidth]{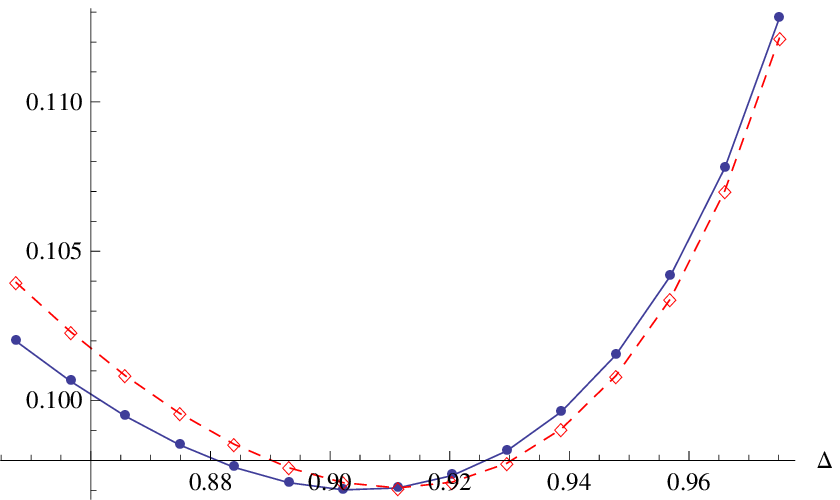}}\\
%%----start of third subfigure----
\subfloat[$r=1.25$]{
\includegraphics[width=0.35\linewidth]{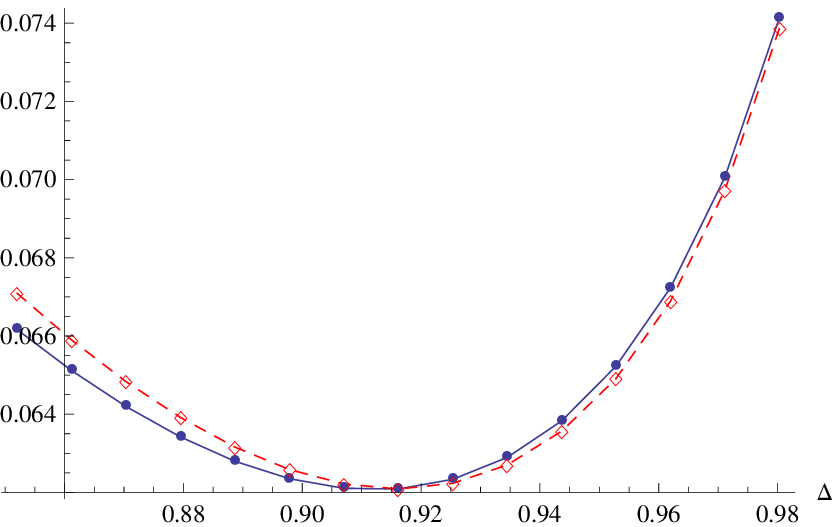}}
\hspace{0.1\linewidth}
%%----start of fourth subfigure----
\subfloat[$r=2.50$]{
\includegraphics[width=0.35\linewidth]{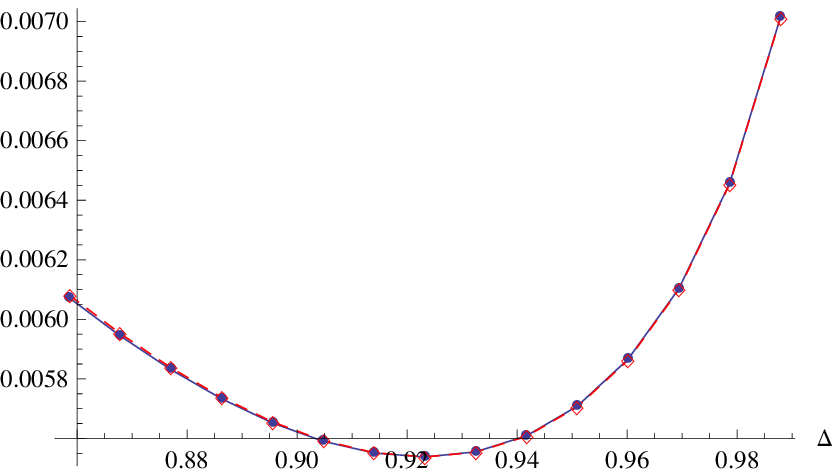}}
\caption{$\mathcal{D}_N$  (full line, filled circle markers) and $(1-F)$ (not to vertical
scale, dashed line, empty diamonds) for a vacuum input state  for a Squeezed Bell-like resource, as a function of $\Delta$ for: (a)
$r=0.75$; (b) $r=1.00$; (c) $r=1.25$; (d) $r=2.50$ . The  central point in each
graphic corresponds to the $\Delta$ that maximizes teleportation fidelity for the input state , and has been made to coincide in both plots}
\label{fig:DFuncvac}  %% label for entire figure
\end{figure}

The vacuum input state is both a coherent state (with poissonian statistics) of displacement $0$,
and a Fock state with average photon number $0$. In figure~(\ref{fig:DFuncvac}), $\mathcal{D}_N$
 and $1-F$  are displayed as functions of the Squeezed
Bell-like resource's $\Delta$, for four values of the two-mode squeezing $r$ thought to go from a
realistic resource up to a resource approximating an \textbf{EPR} state.

It can be seen that the optimal $\Delta$ for $\mathcal{D}_N$ and $1-F$ are noticeably different for
the smaller values of $r$. For higher  values of $r$, they become closer. For very high $r$ the
resource approaches an \textbf{EPR} state, and both optimal $\Delta$s converge to
$\Delta^{opt}_{(2)}$. $\mathcal{D}_N$ and the Frobenius distance (see
equation~(\ref{eq:Frobenius1})) are equal, up to the accuracy with which $\mathcal{D}_N$ has been
calculated. This is in agreement with the results in section~(\ref{CompD:Fidelity}), for input
states that are Fock states.

\subsection{Fock-$1$ Input State}\label{DFuncSamples:Fock1}

\begin{figure}[tb]
\centering
%%----primera----
\subfloat[$r=0.75$]{
\includegraphics[width=0.35\linewidth]{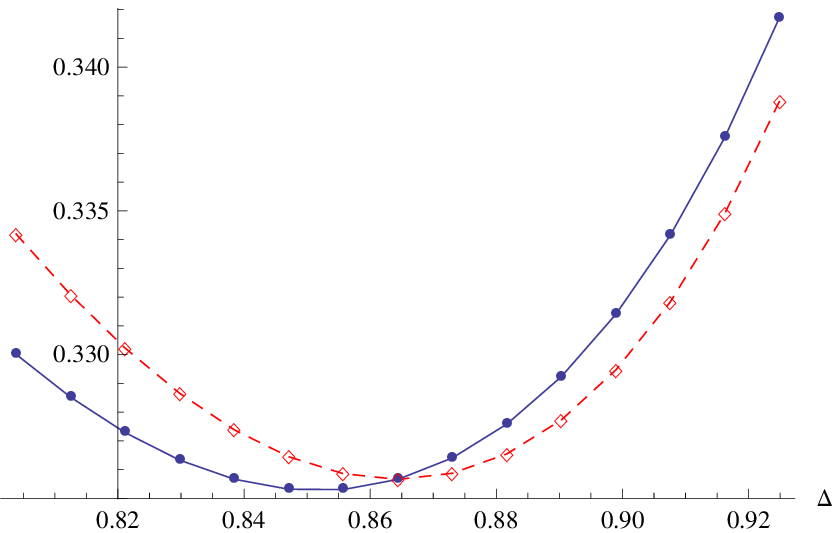}}
\hspace{0.1\linewidth}
%%----start of second subfigure----
\subfloat[$r=1.00$]{
\includegraphics[width=0.35\linewidth]{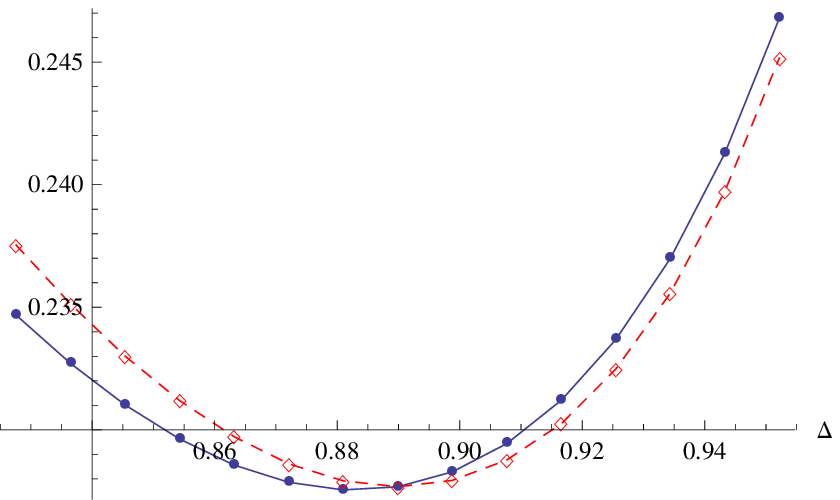}}\\
%%----start of third subfigure----
\subfloat[$r=1.25$]{
\includegraphics[width=0.35\linewidth]{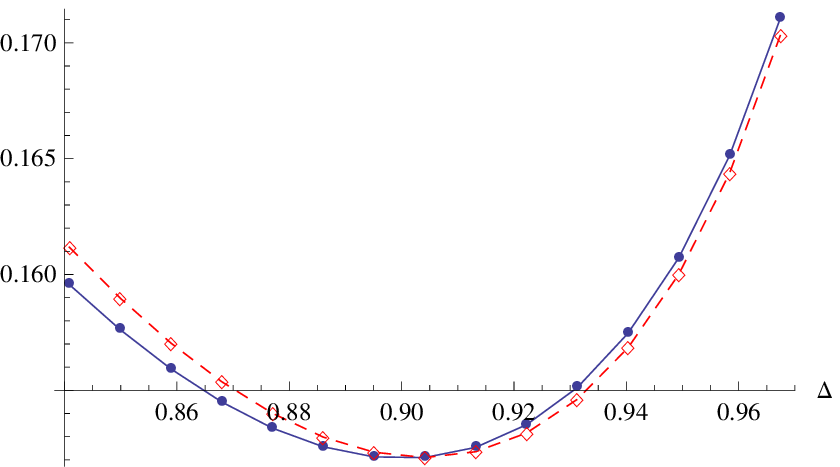}}
\hspace{0.1\linewidth}
%%----start of fourth subfigure----
\subfloat[$r=2.50$]{
\includegraphics[width=0.35\linewidth]{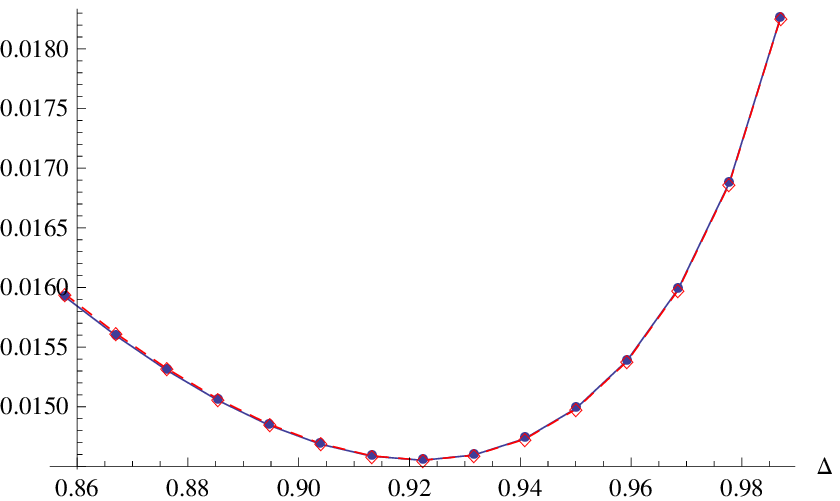}}
\caption{$\mathcal{D}_N$ (full line, filled circle markers) and $(1-F)$ (not to vertical
scale, dashed line, empty diamonds) for a Fock-1 input state for a Squeezed Bell-like resource, as a function  $\Delta$ for: (a)
$r=0.75$; (b) $r=1.00$; (c) $r=1.25$; (d) $r=2.5$ . The central point  in each
graphic corresponds to the $\Delta$ that maximizes teleportation fidelity for the input state and
resource used, and has been made to coincide in both plots}
\label{fig:DFuncf1}  %% label for entire figure
\end{figure}

The Fock-1 ($1$ photon number) input state is a highly non-classical state with sub-poissonian
photon statistics ($g^{(2)}(0)=0$). In figure~(\ref{fig:DFuncf1}),  $\mathcal{D}_N$  and
$1-F$ are displayed.

It can be observed that the optimal $\Delta$ for $\mathcal{D}_N$ and $(1-F)$ are different, at
least for the lower values of $r$. As $r$ grows, the optimal $\Delta$s converge to
$\Delta^{opt}_{(2)}$.

As in the case of the vacuum input state, it occurs that $\mathcal{D}_N$ and the Frobenius distance
(see equation~(\ref{eq:Frobenius1})) are equal, up to the accuracy with which $\mathcal{D}_N$ has been
calculated, for the instances of figure~(\ref{fig:DFuncf1}). This is in agreement with previous
results (see section~(\ref{CompD:Fidelity})) for an input state which is a Fock state.

\begin{figure}[tb]
\centering
%%----primera----
\subfloat[$r=0.75$]{
\includegraphics[width=0.35\linewidth]{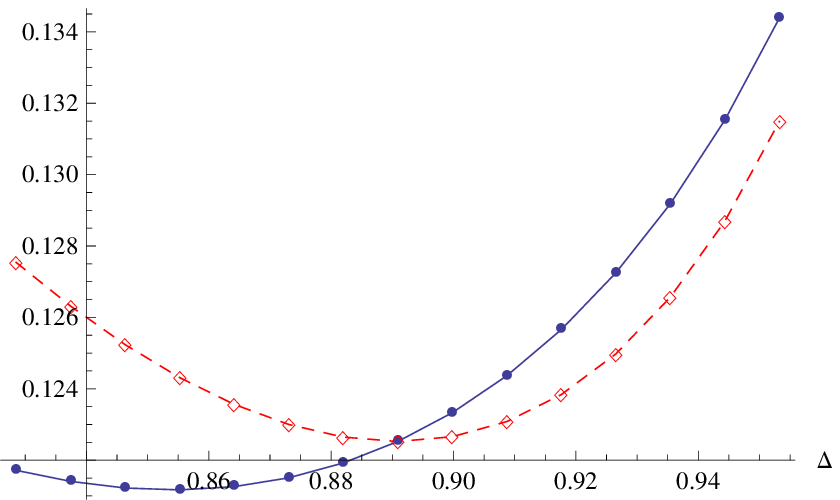}}
\hspace{0.1\linewidth}
%%----start of second subfigure----
\subfloat[$r=1.00$]{
\includegraphics[width=0.35\linewidth]{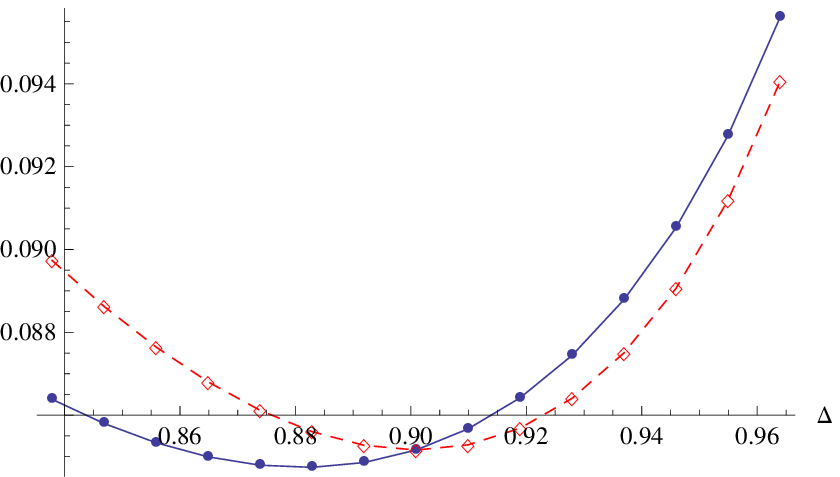}}\\
%%----start of third subfigure----
\subfloat[$r=1.25$]{
\includegraphics[width=0.35\linewidth]{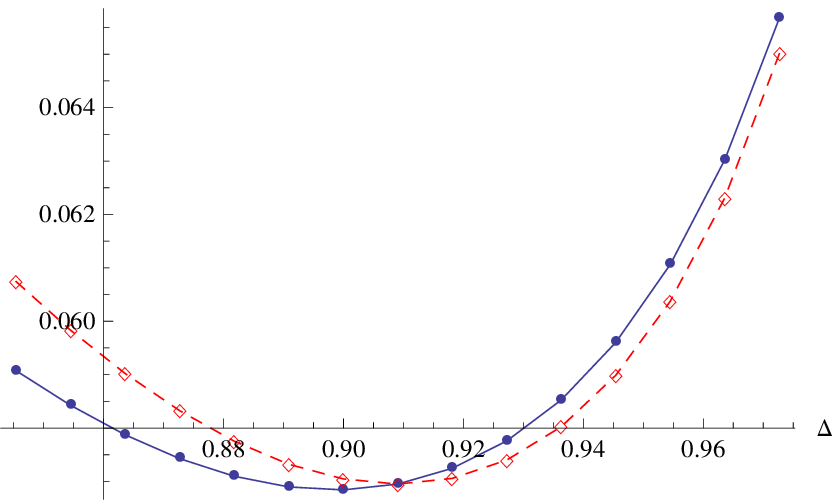}}
\hspace{0.1\linewidth}
%%----start of fourth subfigure----
\subfloat[$r=2.50$]{
\includegraphics[width=0.35\linewidth]{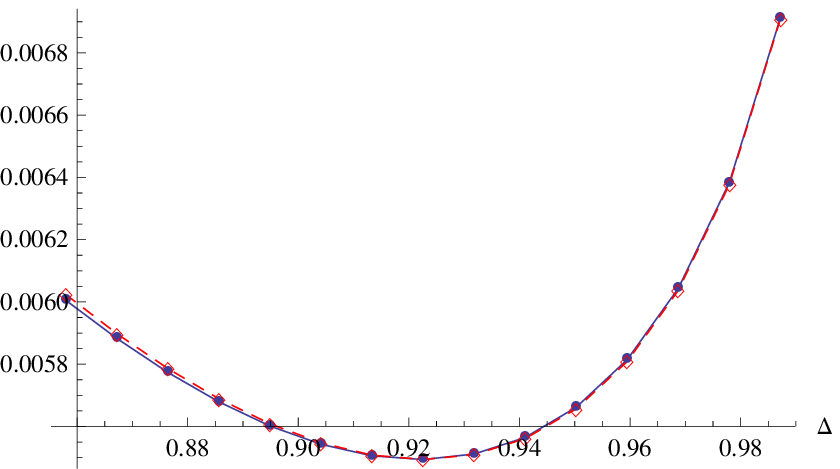}}
\caption{$\mathcal{D}_N$  (full line, filled circle markers) and $(0.5-F)$ (not to
vertical scale, dashed line, empty diamonds) for a mixture of Fock-1 and vacuum input state with
equal probability $P=0.5$ using a Squeezed
Bell-like resource, as a function $\Delta$ for: (a) $r=0.75$; (b) $r=1.00$; (c) $r=1.25$; (d)
$r=2.5$ . The  central point  in each graphic corresponds to the $\Delta$
that maximizes teleportation fidelity for the input state and resource used, and has been made to
coincide in both plots} \label{fig:DFuncmix}   %% label for entire figure
\end{figure}

\subsection{Mixture of Fock-$1$ and Vacuum Input State}\label{DFuncSamples:Mix}

The input state chosen for this example is a mixture of two Fock states, with photon numbers $0$
and $1$, with equal probability $P=0.5$. This
mixture has the characteristic function
\begin{equation}\label{eq:CharMixIn}
    \chi_{mix}(\xi)=0.5\,\chi_{coh,0}(\xi)+0.5\,\chi_{f1}(\xi)
\end{equation}
and is a nonclassical state with sub-poissonian statistics ($g^{(2)}(0)=0$). It is also different
from all the other input states used in this work in that it is a mixed state with a purity
(equation~(\ref{eq:Fidepurity})) equal to $0.5$. The maximum possible fidelity for this input (see
section~(\ref{CompD:Fidelity})) is $0.5$.  $\mathcal{D}_N$  and $0.5-F$ are displayed
in figure~(\ref{fig:DFuncmix}). There is a very noticeable difference, for low $r$, between the
optimal $\Delta$ values for  $\mathcal{D}_N$ and $(0.5-F)$, which is more pronounced than in the
cases of the two Fock states previously studied. This difference becomes smaller for higher $r$, as
expected, and for very high $r$ with the resource approximating an \textbf{EPR} state, there is
convergence to $\Delta^{opt}_{(2)}$.

As in the case of the vacuum and Fock-$1$ input states,  $\mathcal{D}_N$ and the
Frobenius distance are equal, up to the accuracy with which $\mathcal{D}_N$ has been calculated, for
the instances of figure~(\ref{fig:DFuncmix}).

\subsection{Coherent Input State}\label{DFuncSamples:Coherent}

\begin{figure}[tb]
\centering
%%----primera----
\subfloat[$r=0.75$]{
\includegraphics[width=0.35\linewidth]{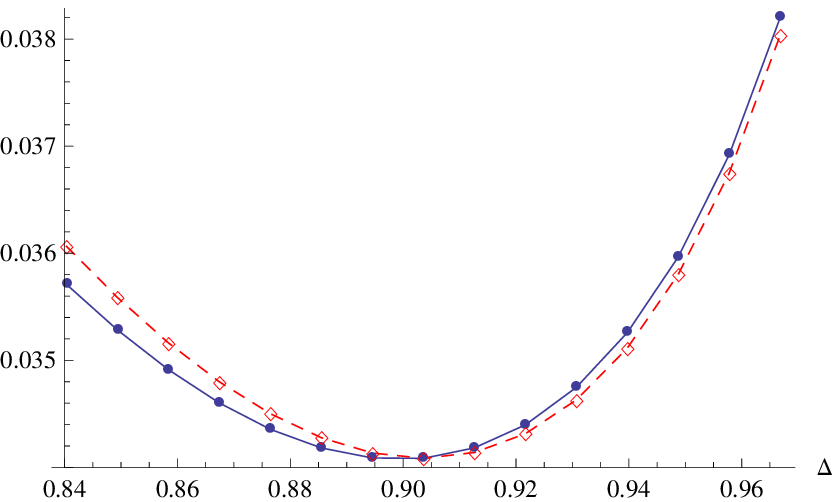}}
\hspace{0.1\linewidth}
%%----start of second subfigure----
\subfloat[$r=1.00$]{
\includegraphics[width=0.35\linewidth]{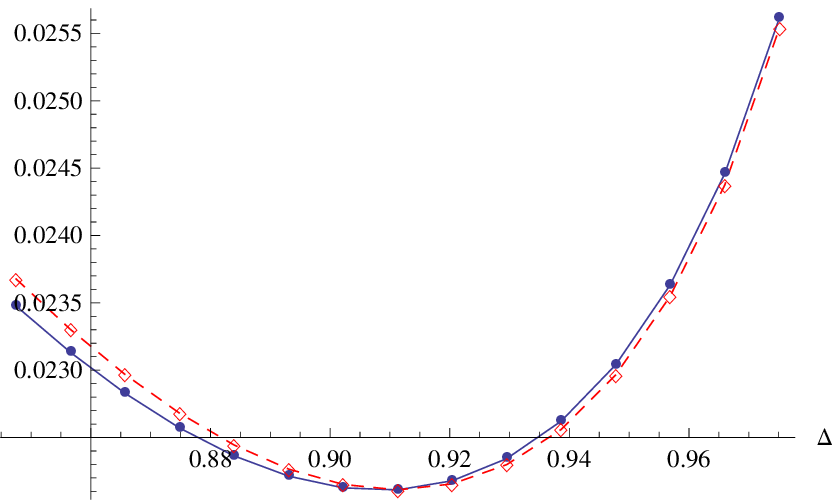}}\\
%%----start of third subfigure----
\subfloat[$r=1.25$]{
\includegraphics[width=0.35\linewidth]{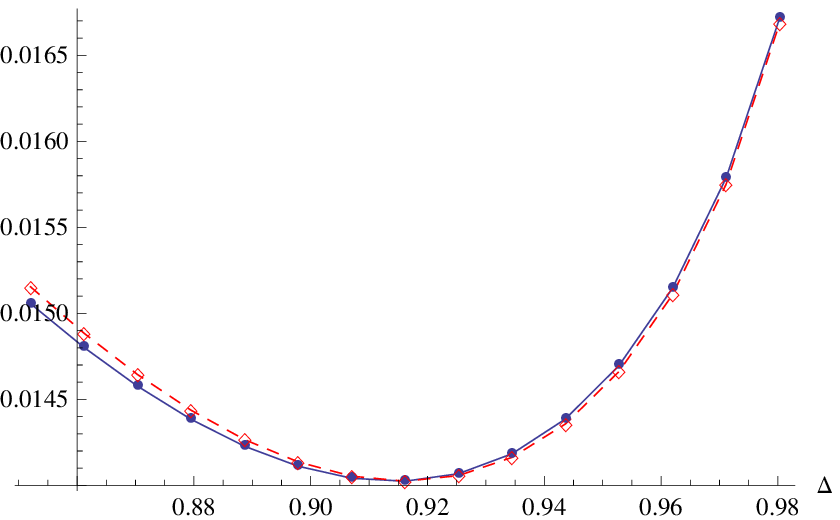}}
\hspace{0.1\linewidth}
%%----start of fourth subfigure----
\subfloat[$r=2.50$]{
\includegraphics[width=0.35\linewidth]{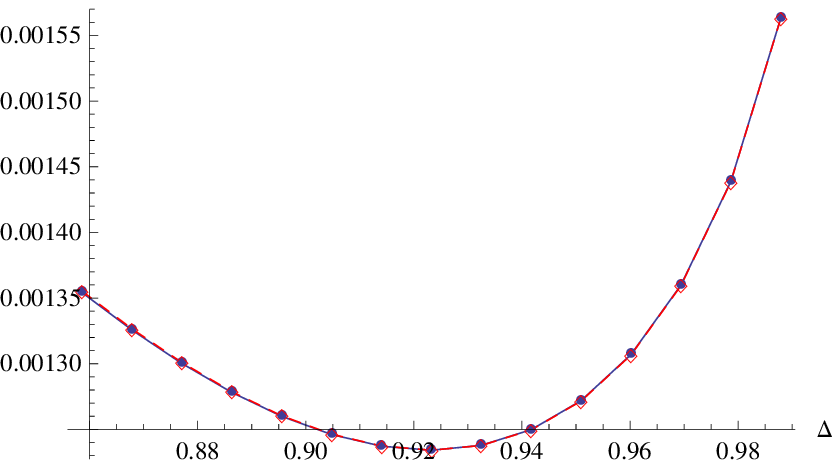}}
\caption{$\mathcal{D}_N$ f (full line, filled circle markers) and $(1-F)$ (not to vertical
scale, dashed line, empty diamonds) for a coherent input state of real displacement
$\beta=2.12928$,  using a Squeezed Bell-like
resource, as a function of the $\Delta$ for: (a) $r=0.75$; (b) $r=1.00$; (c) $r=1.25$; (d) $r=2.5$ .
The  central point  in each graphic corresponds to the $\Delta$ that
maximizes teleportation fidelity for the input state and resource used, and has been made to
coincide for both plots} \label{fig:DFunccoh}    %% label for entire figure
\end{figure}

The coherent input state with displacement $\beta=2.12928$ has poissonian statistics,as
the vacuum input state, and has the same average photon number value ($4.534$) of a
squeezed vacuum with squeezing  $s=1.5$. In figure~(\ref{fig:DFunccoh}),  $\mathcal{D}_N$
 and $(1-F)$ are displayed.

It can be seen that the optimal values of $\Delta$ for $\mathcal{D}_N$ and $1-F$ coincide for the
coherent input state, for the values of $r$  displayed. This is an unexpected result for an input
state that is neither a Fock state, nor a mixture of Fock states. The shared minimum does converge,
for higher $r$, to $\Delta^{opt}_{(2)}$.

\begin{figure}[tb]
\centering
%%----primera----
\subfloat[$r=0.75$]{
\includegraphics[width=0.35\linewidth]{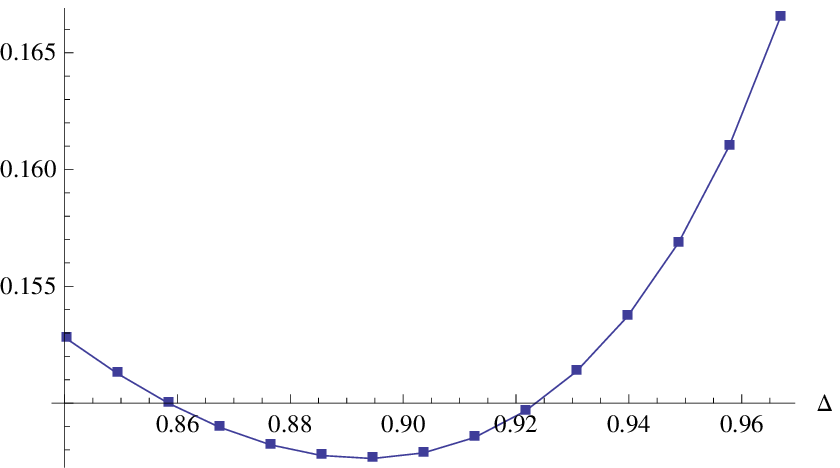}}
\hspace{0.1\linewidth}
%%----start of second subfigure----
\subfloat[$r=1.00$]{
\includegraphics[width=0.35\linewidth]{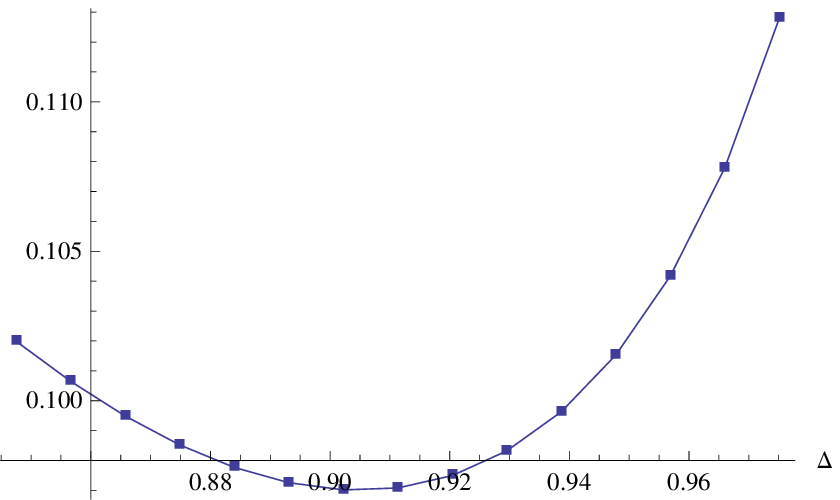}}\\
%%----start of third subfigure----
\subfloat[$r=1.25$]{
\includegraphics[width=0.35\linewidth]{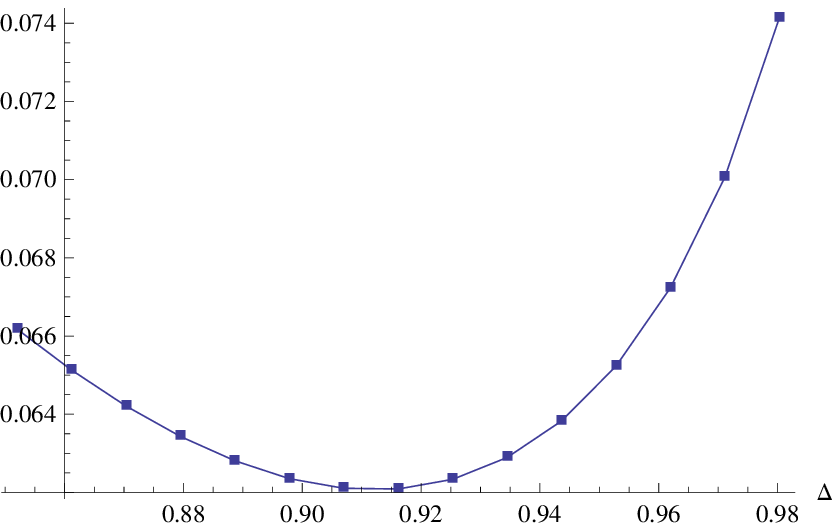}}
\hspace{0.1\linewidth}
%%----start of fourth subfigure----
\subfloat[$r=2.50$]{
\includegraphics[width=0.35\linewidth]{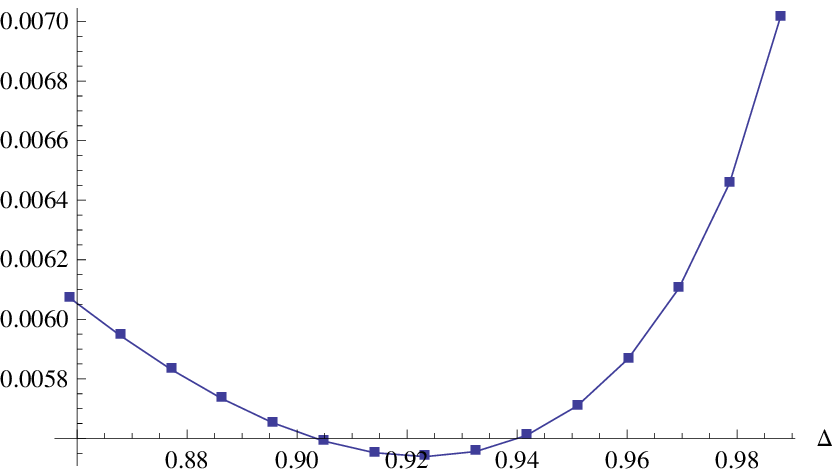}}
\caption{Frobenius distance $\delta_{F}$  for a coherent input state of real
displacement $\beta=2.12928$,  using a Squeezed
Bell-like resource, as a function of  $\Delta$ for: (a) $r=0.75$; (b) $r=1.00$; (c) $r=1.25$; (d)
$r=2.5$ . The central point  in each graphic corresponds to the $\Delta$
that maximizes teleportation fidelity for the input state and resource used} \label{fig:Frobcoh}     %% label for entire figure
\end{figure}

The Frobenius distance between the input coherent state and the teleportation output is shown in
figure~(\ref{fig:Frobcoh}). $\mathcal{D}_N$ is not equal the Frobenius distance
$\delta_{F}$, for the coherent input state. However, the optimal $\Delta$ value for the Frobenius
distance coincides with that for $\mathcal{D}_N$, for all the $r$ values used.

\subsection{Squeezed Vacuum Input State}\label{DFuncSamples:Squeezed}

\begin{figure}[tb]
\centering
%%----primera----
\subfloat[$r=0.75$]{
\includegraphics[width=0.35\linewidth]{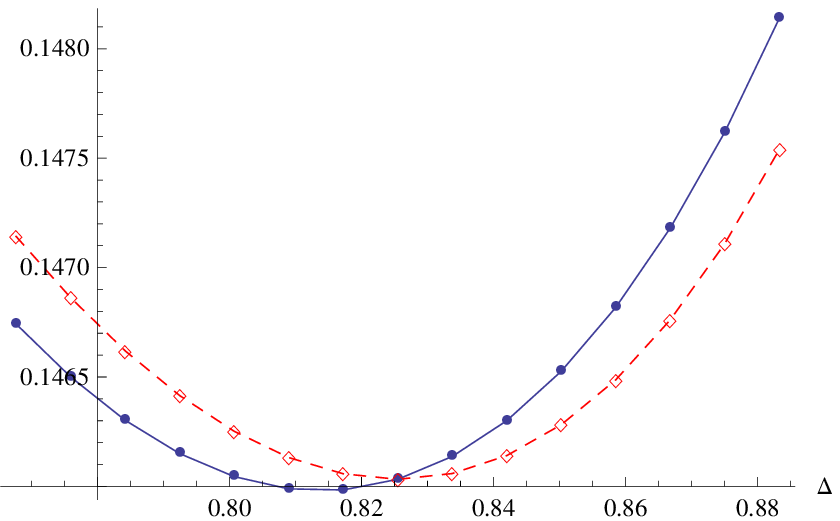}}
\hspace{0.1\linewidth}
%%----start of second subfigure----
\subfloat[$r=1.00$]{
\includegraphics[width=0.35\linewidth]{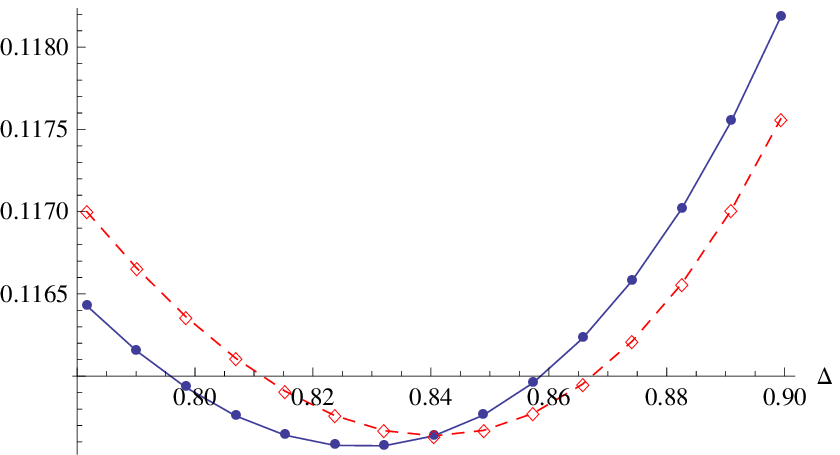}}\\
%%----start of third subfigure----
\subfloat[$r=1.25$]{
\includegraphics[width=0.35\linewidth]{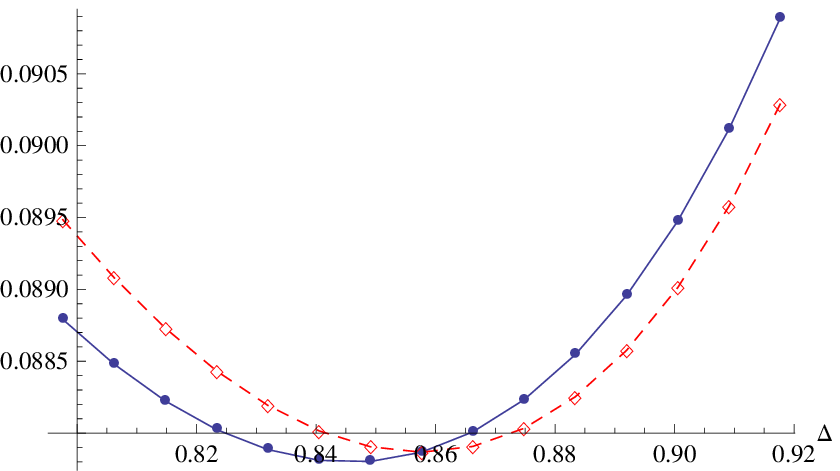}}
\hspace{0.1\linewidth}
%%----start of fourth subfigure----
\subfloat[$r=2.50$]{
\includegraphics[width=0.35\linewidth]{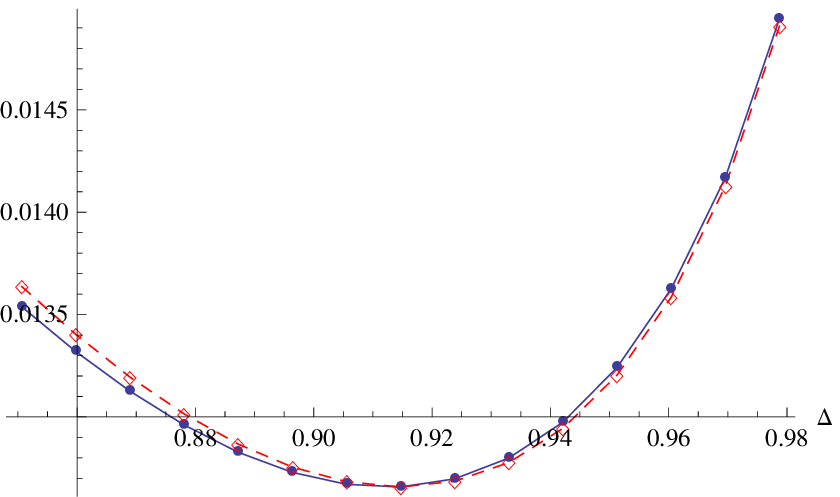}}
\caption{$\mathcal{D}_N$  (full line, filled circle markers) and $(1-F)$ (not to vertical
scale, dashed line, empty diamonds) for a squeezed vacuum (squeezing $s=1.5$) input state,  using a Squeezed Bell-like resource, as a function of the
$\Delta$ parameter of the Squeezed Bell-like resource for: (a) $r=0.75$; (b) $r=1.00$; (c) $r=1.25$; (d) $r=2.5$ . The central point  in each graphic corresponds to the $\Delta$ that maximizes teleportation fidelity for
the input state and resource used, and has been made to coincide for both plots}
\label{fig:Dfuncsqv}       %% label for entire figure
\end{figure}

The last input state we will consider is the squeezed vacuum input state with $s=1.5$. It is a highly non-classical
state with sub-poissonian photon statistics, and  has an average photon number equal to that of the
coherent state used above.  $\mathcal{D}_N$ and  $(1-F)$  are displayed
in figure~(\ref{fig:Dfuncsqv}).

The results for this input state are very different from those of both the vacuum state or the
coherent state, for all but the highest values of $r$. There is considerable distance between the
optimal $\Delta$s for $\mathcal{D}_N$ and $(1-F)$, the greatest difference among all the input
states used. The expected convergence of $\Delta$, to the value $\Delta^{opt}_{(2)}$ with
increasing $r$ is much slower than for the other input states. This is the signature of a highly
non-classical input state that is no Fock state, nor a mixture of Fock states.

The Frobenius distance between squeezed vacuum input and teleportation output is displayed in
figure~(\ref{fig:Frobcoh}). It can be seen that  $\mathcal{D}_N$  is not equal to the Frobenius
distance $\delta_{F}$. The optimal $\Delta$s for the Frobenius distance, $\mathcal{D}_N$ and
$(1-F)$ are all different. Only for high $r$ when the resources approach the \textbf{EPR} state, do
the optimal $\Delta$s converge for the three functionals to $\Delta^{opt}_{(2)}$.

\begin{figure}[tb]
\centering
%%----primera----
\subfloat[$r=0.75$]{
\includegraphics[width=0.35\linewidth]{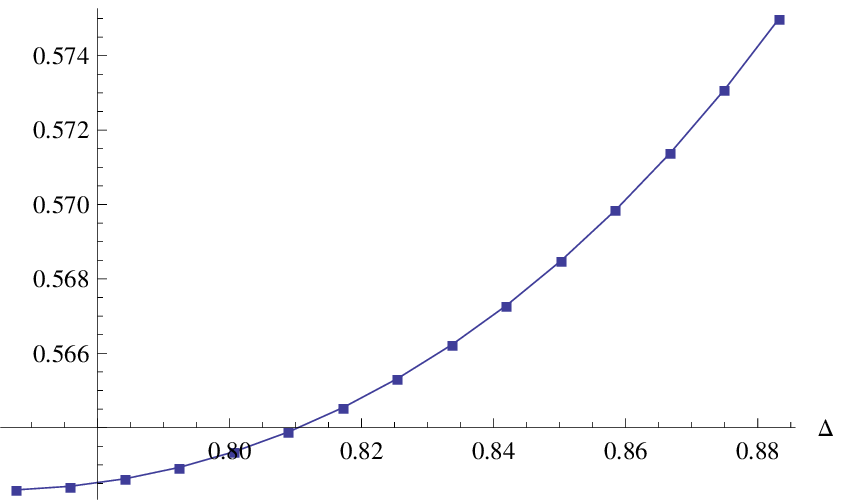}}
\hspace{0.1\linewidth}
%%----start of second subfigure----
\subfloat[$r=1.00$]{
\includegraphics[width=0.35\linewidth]{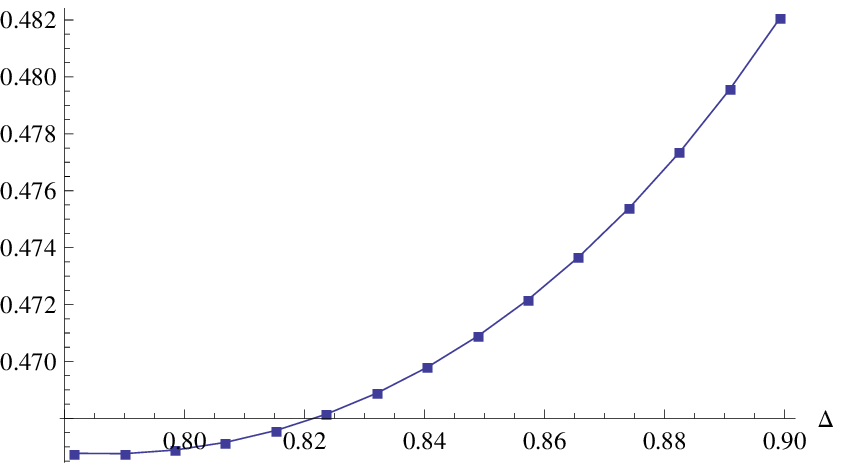}}\\
%%----start of third subfigure----
\subfloat[$r=1.25$]{
\includegraphics[width=0.35\linewidth]{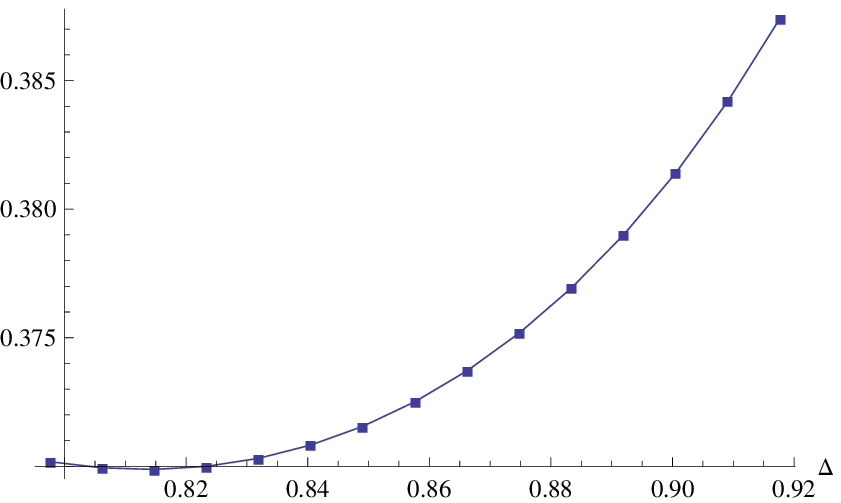}}
\hspace{0.1\linewidth}
%%----start of fourth subfigure----
\subfloat[$r=2.50$]{
\includegraphics[width=0.35\linewidth]{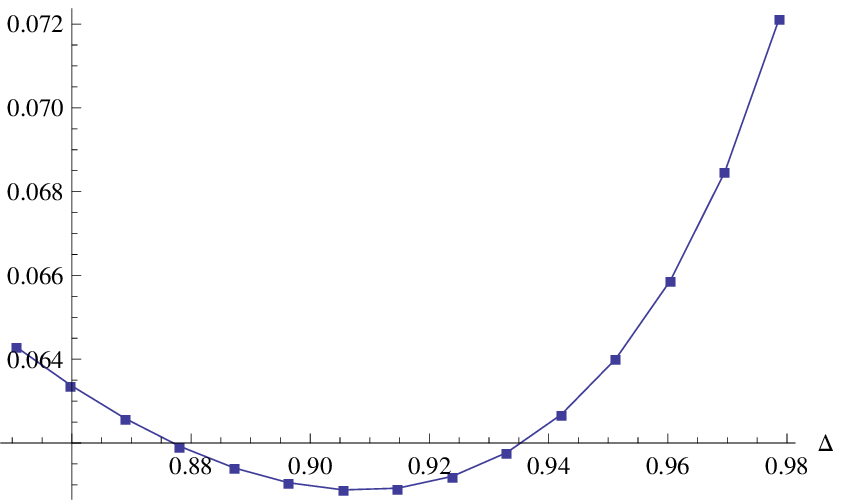}}
\caption{Frobenius distance $\delta_{F}$ for a squeezed vacuum input state of
squeezing $s=1.5$, and the output of teleportation of such an input using a Squeezed Bell-like
resource, as a function of the $\Delta$ parameter of the Squeezed Bell-like resource. The two-mode
squeezing $r$ of the resource state is equal to: (a) $0.75$; (b) $1.00$; (c) $1.25$; (d) $2.5$ .
The central point in each graphic corresponds to the $\Delta$ that
maximizes teleportation fidelity for the input state and resource used} \label{fig:Frobsqv}     %% label for entire figure
\end{figure}

\clearpage %Pone las figuras sobrantes ANTES que la discusion y la bibliografia.

\section{Discussion}\label{Discuss}

In this paper we have considered the distortions that teleportation induces on second and higher
order momenta of gaussian and non gaussian states. Working within the characteristic function
representation of \textbf{CV} quantum teleportation we derive expressions for average values of
output state observables in terms of the input state and the resource of teleportation. The
structure of the output characteristic function as a product between the input characteristic
function and the transfer function and as the Fourier transform of a convolution of the respective
pseudo-distributions leads us to recognize the importance of cumulants as the particular functions
of raw momenta which are additive with contributions of the input and of the transfer functions.
For cumulants distortions produced by teleportation are universal in the sense that for any input,
they only depend on the resource properties.

We  explore  teleportation efficiency for cumulants and principal momenta  using Squeezed-Bell like
states as the resource. These states depend on a free parameter $\Delta$ and for high squeezing $r$
approximate ideal $EPR$ resources. We obtain optimal resources for the different functions and
different input states and compare with the optimal fidelity resources discussed in
Ref.~\cite{TelepNonGauss}.

For second order momenta which are also second order cumulants we determine from the transfer
function of the resource the optimal  value $\Delta^{opt}_{(2)}=\frac{\sqrt{2+\sqrt{2}}}{2}$
independently of  the squeezing. For fourth order cumulants we determine numerically the different
optimal value $  \Delta_{(4)}^{opt}\approx 0.985294$. We also determine the optimal resources for
teleportation of the fourth order central moment for three different input states and show that
they depend on input state and on the squeezing.

For high squeezing we show that the optimal fidelity resources which also depend on the input
properties tend to the resource with $\Delta^{opt}_{(2)}$. This is explained by noting that
distortions associated with the higher order momenta decrease faster than those associated with the
second order moment when $r$ becomes large. Even for fourth order momenta which are not cumulants
we show that the optimal resources for the different input states tend to the one with
$\Delta^{opt}_{(2)}$.

We consider the functional $\mathcal{D}_N$ of  quadratic deviations and compare photon
statistics between input and output states.We show that it  approximates Frobenius distance for
input states that are mixtures of Fock states. Minimizing  $\mathcal{D}_N$  allows to determine numerically optimal values of $\Delta$ for various
inputs. They  differ from the corresponding maximum fidelity values. As expected, $\mathcal{D}_N$
approximates Frobenius distance better for Fock states or mixtures of Fock states. For every input used and for $r$ arbitrarily large, Frobenius distance, Fidelity and
$\mathcal{D}_N$ optimal values of $\Delta$ converge to $\Delta^{opt}_{(2)}$.

We note that it is straightforward to devise functionals analogous to $\mathcal{D}_N$, using other
orthonormal, complete bases of eigenstates. These functionals will be good approximations to the
Frobenius distance for mixtures of states of the basis selected.

Our results support the view that optimal fidelity resources are in first approximation well suited
also for an optimal teleportation of more specific characteristics of the input but that in a realistic
situation with $r$ not to large there is room for improvement of the resource.

\bibliographystyle{unsrt}

\end{document}